\documentclass[a4paper,twocolumn,11pt,accepted=2025-05-12]{quantumarticle}
\pdfoutput=1
\usepackage[utf8]{inputenc}
\usepackage[english]{babel}
\usepackage[T1]{fontenc}
\usepackage{amsmath}
\usepackage[unicode]{hyperref}
\usepackage[numbers,sort&compress]{natbib}

\def\Ket#1{\left|#1\right>}

\begin{document}

\title{Transfer and routing of Gaussian states through quantum complex networks with and without community structure}

\author{Markku Hahto}
\affiliation{Department of Physics and Astronomy, University of Turku, FI-20014, Turun Yliopisto, Finland}
\orcid{0009-0006-2838-4626}

\author{Johannes Nokkala}
\affiliation{Department of Physics and Astronomy, University of Turku, FI-20014, Turun Yliopisto, Finland}
\orcid{0000-0002-5052-9813}

\author{Guillermo Garc{\'i}a-P{\'e}rez}
\affiliation{Algorithmiq Ltd, Kanavakatu 3 C, FI-00160, Helsinki, Finland}
\orcid{0000-0002-9006-060X}
 
\author{Sabrina Maniscalco}
\affiliation{Algorithmiq Ltd, Kanavakatu 3 C, FI-00160, Helsinki, Finland}
\affiliation{Department of Physics, University of Helsinki, FI-00014 Helsinki, Finland}
\orcid{0000-0001-8559-0828}

\author{Jyrki Piilo}
\affiliation{Department of Physics and Astronomy, University of Turku, FI-20014, Turun Yliopisto, Finland}
\orcid{0000-0002-5595-873X}

\maketitle

\begin{abstract}
The goal in quantum state transfer is to avoid the need to physically transport carriers of quantum information. This is achieved by using a suitably engineered Hamiltonian that induces the transfer of the state of one subsystem to another. A less known generalization of state transfer considers multiple systems such that any pair can exchange quantum information and transfers can take place at any time, starting and stopping independently. This is sometimes called routing of quantum states. State transfer in particular has received a great deal of attention, however the vast majority of results in both state transfer and routing concern qubits transferred in a network of restricted structure. Here we consider routing of single-mode Gaussian states and entanglement through complex networks of quantum harmonic oscillators. We compare a protocol where the transfer is completed in a single step but the effective Hamiltonian only approximately transfers the state with one where the transfer can in principle be perfect but the transfer is done in two steps, and also illustrate the state-dependency of the transfer fidelity with paradigmatic Gaussian states as well as number states. We find that even in a random and homogeneous network, the transfer fidelity still depends on the degree of the nodes for any link density, and that in both random and complex networks it is the community structure that controls the appearance of higher frequency normal modes useful for transfer. Finally, we find that networks of sufficient complexity may have superior routing performance over superficially similar random networks. Our results pave the way for further exploration of the role of community structure in state transfer and related tasks.
\end{abstract}

\section{\label{sec:introduction}Introduction}

Quantum state transfer is the task of transferring a state from some initial system to a target system through Hamiltonian dynamics of a network of many interacting systems. This could be used to realize high fidelity channels connecting different quantum processors, perhaps eventually leading to all solid state quantum information processing on a chip \cite{bose2003quantum,nikolopoulos2014quantum,lewis2023low}. State transfer is also considered from a more formal point of view as a fundamental problem in mathematical physics \cite{kay2011basics,godsil2012state}. Sometimes networks with special symmetries that can naturally transfer the state are sought. This can potentially lead to a fast transfer with minimal control requirements, though heavy restrictions are imposed on the network \cite{christandl2005perfect,kostak2007perfect,kay2011basics,portes2013perfect}. Alternatively, the network may be given and one tries to achieve transfer assuming some limited control over a small part of it, typically by engineering a situation where the sender and receiver effectively interact with only a single normal mode of the rest of the network which then acts as a communication channel \cite{plenio2005high,wojcik2007multiuser,paganelli2013routing,nicacio2016coupled}. In this approach the standard choice is the normal mode corresponding to the smallest eigenvalue of the network, since in the case that the Hamiltonian is proportional to the Laplace matrix of the network, the mode has by construction uniform overlap with all the nodes \cite{merris1998laplacian}, and for example in Erd{\H o}s-R{\'e}nyi networks it is well isolated from the other normal modes due to a large spectral gap \cite{jamakovic2008robustness} which has been shown to be relevant to state transfer \cite{plenio2005high}. Both approaches have been used in fermionic systems, often to transfer single qubits \cite{christandl2005perfect,jonckheere2015information,alsulami2022unitary}, and in bosonic systems, notably to transfer of entanglement \cite{plenio2004dynamics,plenio2005high,chudzicki2010parallel}, number and coherent states \cite{portes2013perfect} or excitations \cite{brown2011coupled}. In the continuous variable case state transfer is of particular interest when distances are short since unlike with qubits, perfect teleportation is impossible \cite{braunstein1998teleportation,pirandola2006quantum}. In general, it can be thought of as a two-user protocol where the users are the sender and the receiver of the state to be transferred. Logically, this corresponds to a two-port bus connecting the sender and the receiver. In particular, even if the users can be freely chosen, only a single transfer takes place at any given time. 

We consider a multi-user generalization of state transfer called quantum state routing \cite{paganelli2013routing}. Here, any pair of users can communicate and transfers may happen at any time such that they can be initiated and completed independently. This realizes a bus capable of receiving and forwarding quantum information continuously with multiple active ports. Note that this should not be confused with parallel state transfer \cite{chudzicki2010parallel} where transfers have fixed destinations and cannot occur independently. Also other tasks are sometimes called routing \cite{sadlier2020state,webber2020efficient,bapat2021quantum,sinha2022qubit} --  notably directing a single or multi-qubit or qudit state to indicated nodes with a coined quantum walk \cite{zhan2014perfect,li2021discrete,gao2023demonstration} or realizing a controllable three-way junction with a chiral walk \cite{bottarelli2023quantum,palaiodimopoulos2023chiral}. So far, routing has been considered in fermionic systems where external users couple to a closed \cite{wojcik2007multiuser,paganelli2013routing} or open \cite{yousefjani2020simultaneous} chain of spins and exchange single qubits over its normal modes --  and within a network with special direct sum structure \cite{pemberton2011perfect} -- which decomposes into non-interacting short spin chains playing a role somewhat similar to normal modes. A bosonic ring was considered in \cite{plenio2005high} to transfer Gaussian entanglement between external users, but limited to sequential case since only one normal mode was used. Although the special networks considered previously facilitate an analytic treatment and may lead to a favorable normal mode structure, the case of complex networks remains uncharted. In the case of using only the slowest normal mode, does it matter where users couple? What properties control the appearance of multiple useful normal modes required to achieve multiple independent transfers? Do networks of genuine complexity have subtle features giving them an edge over random networks?

We address these questions by considering transfer and routing of Gaussian states and entanglement in both random and empirical bosonic networks. We compare and contrast two protocols, observing also how a resource such as squeezing or entanglement may be transferred quite well even if phase differences cause the fidelity to be relatively low. We also confirm the applicability of our scheme to non-Gaussian states by demonstrating high transfer fidelities for number states. We show how even in Erd{\H o}s-R{\'e}nyi networks---considered largely featureless as networks---the external users are not equal. Specifically the degree of the node the user couples to correlates with the fidelity independently of the link density. This makes an abundance of high degree nodes beneficial overall, and therefore also the degree distribution relevant to the transfer fidelity. 

\begin{figure}[t]
\centering
\includegraphics[trim=0cm 11cm 7.5cm 0cm,clip=true,width=0.48\textwidth]{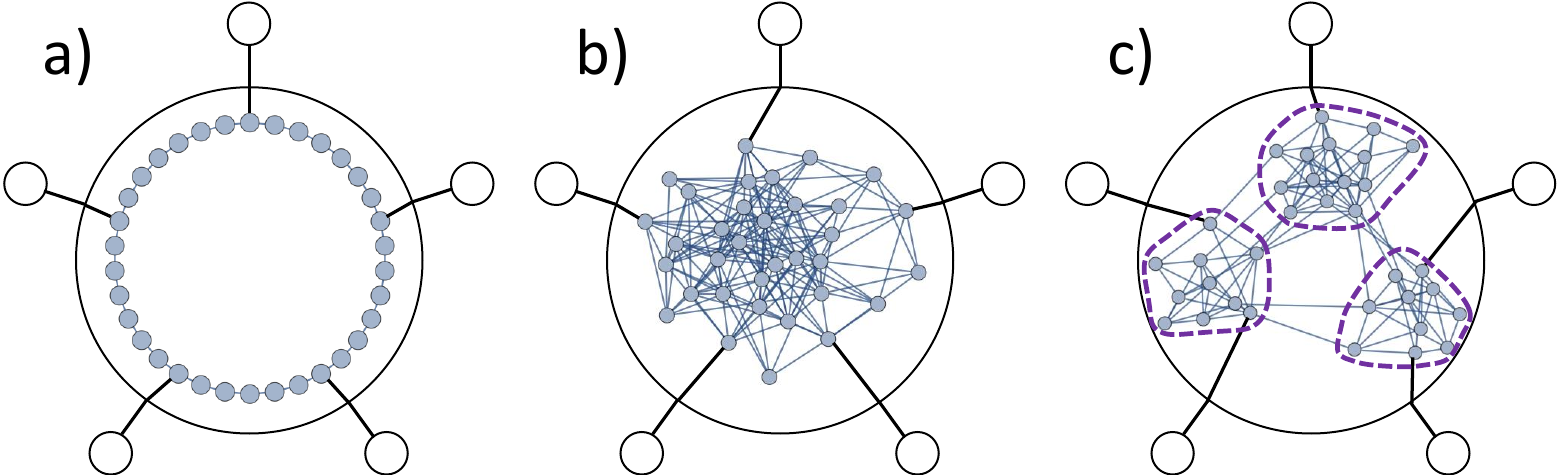}
\caption{\label{fig:routing} The routing problem and some different approaches: $a)$ a ring, $b)$ a random network with homogeneous link density and $c)$ a network with a community structure, with the three communities circled with dashed line. The white circles are users that can act as both senders and receivers of quantum information. Enclosed in the big circle is a central system that should facilitate routing such that its Hamiltonian (consisting of subsystems and interaction terms indicated by the blue circles and lines, respectively) can remain fixed and there is no need to control its state. Only local control of the users and their couplings to the central system, indicated by the black lines, should suffice to facilitate independent and possibly overlapping transfers between any pairs of users.}
\end{figure}

We then reveal the crucial role of community structure in facilitating routing in complex networks, a network property that has formerly remained mostly unexplored also in state transfer. Community structure, an important property in many real networks, is characterized by a non-uniform link density \cite{girvan2002community}. It breaks the network into densely connected subnetworks---the communities---with substantially fewer links connecting nodes belonging to different communities. This is exemplified in Fig.~\ref{fig:routing} \textit{c)} where the communities are circled. Since no formal definition exists, there is no canonical way to find the communities of a network. How well a given division does may be quantified through different measures such as modularity \cite{newman2004finding,fortunato2010community}. Considering a social network with a known community structure, we further identify which nodes allow high fidelity transfer over given normal modes to reveal the finer structure of the available communication channels. This also finds a finer community structure with high modularity as a by-product. Finally, we introduce tentative measures for routing capacity and find that a sufficiently complex network may outperform a random network with superficially similar features.

The rest of this work is structured as follows. In Sec.~\ref{sec:routing} we introduce the model and the scheme for using it to route single-mode Gaussian states and entanglement. We also compare the transfer performance and speed of the one and two step transfer protocols, and how the former depends on the chosen state and the used figure of merit. In Sec.~\ref{sec:random} we consider random networks both without and with a community structure, revealing its role in the routing capabilities of the network. Similar results are also found for some empirical networks, which we consider in Sec.~\ref{sec:empirical}. Here we also benchmark the empirical networks to their randomized counterparts. We conclude in Sec.~\ref{sec:conclusions} where we also discuss new questions our results have raised.

\section{Quantum state routing}\label{sec:routing}

\subsection{The model\label{sec:model}}

We consider a scheme where a network of interacting quantum systems mediates state transfer between external systems coupled to it. The network Hamiltonian is static and its state need not be controlled; transfers are carried out by only local control of the external systems and the coupling terms to the network. When any pair of external systems can exchange quantum information and multiple transfers may start and stop independently we speak of routing, see Fig.~\ref{fig:routing}.

The network consists of $N$ identical quantum harmonic oscillators interacting with springlike couplings of constant magnitude $g$. Such units are used that the reduced Planck constant $\hbar=1$ and the Boltzmann constant $k_B=1$. Arbitrary units are used for coupling strengths and frequencies. All oscillators have unit mass and a bare frequency $\omega_0$. The oscillators have position and momentum operators $q_j=(a_j^\dagger+a_j)/\sqrt{2\omega_0}$, $p_j=\mathrm{i}\sqrt{\omega_0/2}(a_j^\dagger-a_j)$ where $a_j$ ($a_j^\dagger$) is the  annihilation (creation) operator of oscillator $j$ satisfying the commutation relations $[a_i,a_j^\dagger]=\delta_{ij}$.

Let $\mathbf{q}^\top=\{q_1,q_2,\ldots,q_N\}$ and $\mathbf{p}^\top=\{p_1,p_2,\ldots,p_N\}$ be the vectors of position and momentum operators of the network oscillators. Now the network Hamiltonian can be conveniently expressed as
\begin{equation}
    H=\frac{\mathbf{p}^\top\mathbf{p}}{2}+\frac{\mathbf{q}^\top(\omega_0^2\mathbf{I}+g\mathbf{L})\mathbf{q}}{2},
    \label{eq:Hamiltonian}
\end{equation}
where $\mathbf{I}$ is the identity matrix and $\mathbf{L}$ is the network Laplace matrix, which has diagonal elements $\mathbf{L}_{ii}=d_i$ where $d_i$ is the degree of oscillator $i$, or the number of directly coupled other oscillators, whereas the off-diagonal elements $\mathbf{L}_{ij}=-1$ if the oscillators $i$ and $j$ are coupled and $0$ otherwise. If the degrees are collected into a diagonal matrix $\mathbf{D}$, then the Laplace matrix can be expressed in terms of the network adjacency matrix $\mathbf{V}$ as $\mathbf{L}=\mathbf{D}-\mathbf{V}$. The Hamiltonian is by construction diagonalizable to an equivalent diagonal form of non-interacting normal modes \cite{tikochinsky1979diagonalization}.

The external systems are likewise quantum harmonic oscillators that each couple to a randomly selected network oscillator with a linear coupling term of the form $H_I=-kq_Sq_i$ where $q_S$ ($q_i$) is the position operator of the external system (network oscillator). It is assumed that both the coupling strength $k$ and the system frequency $\omega_S$ can be tuned freely. Only one external system may be coupled to each network oscillator as otherwise the systems would be physically close and an indirect transfer over the network would not be necessary. For similar reasons a given external system cannot switch the network oscillator it is coupled to, although it can decouple from the network by setting $k=0$.

We consider the transfer of Gaussian states \cite{ferraro2005gaussian,adesso2014continuous} which are completely determined by their covariance matrix and first moments vector. Let $\mathbf{x}^\top=\{q_1,q_2,\ldots,p_1,p_2,\ldots\}$. Then the covariance matrix has elements $\sigma(\mathbf{x})_{ij}=\langle\mathbf{x}_i\mathbf{x}_j+\mathbf{x}_j\mathbf{x}_i\rangle/2-\langle\mathbf{x}_i\rangle\langle\mathbf{x}_j\rangle$ whereas the first moments vector is simply $\langle\mathbf{x}\rangle_i=\langle\mathbf{x}_i\rangle$. In particular we consider the transfer of squeezed vacuum states $\Ket{r,\varphi}$ with squeezing parameter $r$ and phase $\varphi$, and coherent states $\Ket{\alpha}$ with complex displacement $\alpha$. To study the transfer of entanglement, we use as a sender two mode squeezed vacuum states $\Ket{s,\phi}$ with two mode squeezing parameter $s$ and phase $\phi$ while one of the modes of the bipartite system interacts with the network. The covariance matrices and vectors of means for all of the above mentioned states are given in App.~\ref{app:covariancesandvectors}. Although these are pure states the marginal states of $\Ket{s,\phi}$ are mixed when $s>0$ because then the state is entangled. The initial states for the receiver and the network can in principle be arbitrary. In the numerical simulations the receiver is initially in a vacuum state $\Ket{0}$, and the networks is either in a product state of vacuums $\Ket{0}^{\otimes N}$ (Sec.~\ref{sec:erdosrenyi} and \ref{sec:empirical}A) or the ground state of $H$ (Sec.~\ref{sec:empirical}B).

As figures of merit we primarily use the (Uhlmann) fidelity, defined to be
\begin{equation}
    \mathcal{F}(\rho_1,\rho_2)=\left(\mathrm{Tr}\sqrt{\sqrt{\rho_1}\rho_2\sqrt{\rho_1}}\right)^2
    \label{eq:fidelity}
\end{equation}
where $\rho_1$ and $\rho_2$ are density operators of the initial and transferred state. We will also use efficiency,
\begin{equation}
    \mathrm{eff}_X(t) = \frac{X_R(t)}{X_S(0)}
    \label{eq:efficiency}
\end{equation}
which is the ratio of a specific resource $X$ that was transferred to the receiver $R$ at time $t$ to the initial amount at the sender $S$. The resource $X$ can be for example squeezing or entanglement. For single mode Gaussian states considered here the fidelity can be easily calculated directly from the covariance matrices and first moment vectors.

\subsection{State transfer over the normal modes\label{sec:shortchains}}

\begin{figure}[ht]
\centering
\includegraphics[trim=0cm 0cm 0.0cm 0cm,clip=true,width=0.48\textwidth]{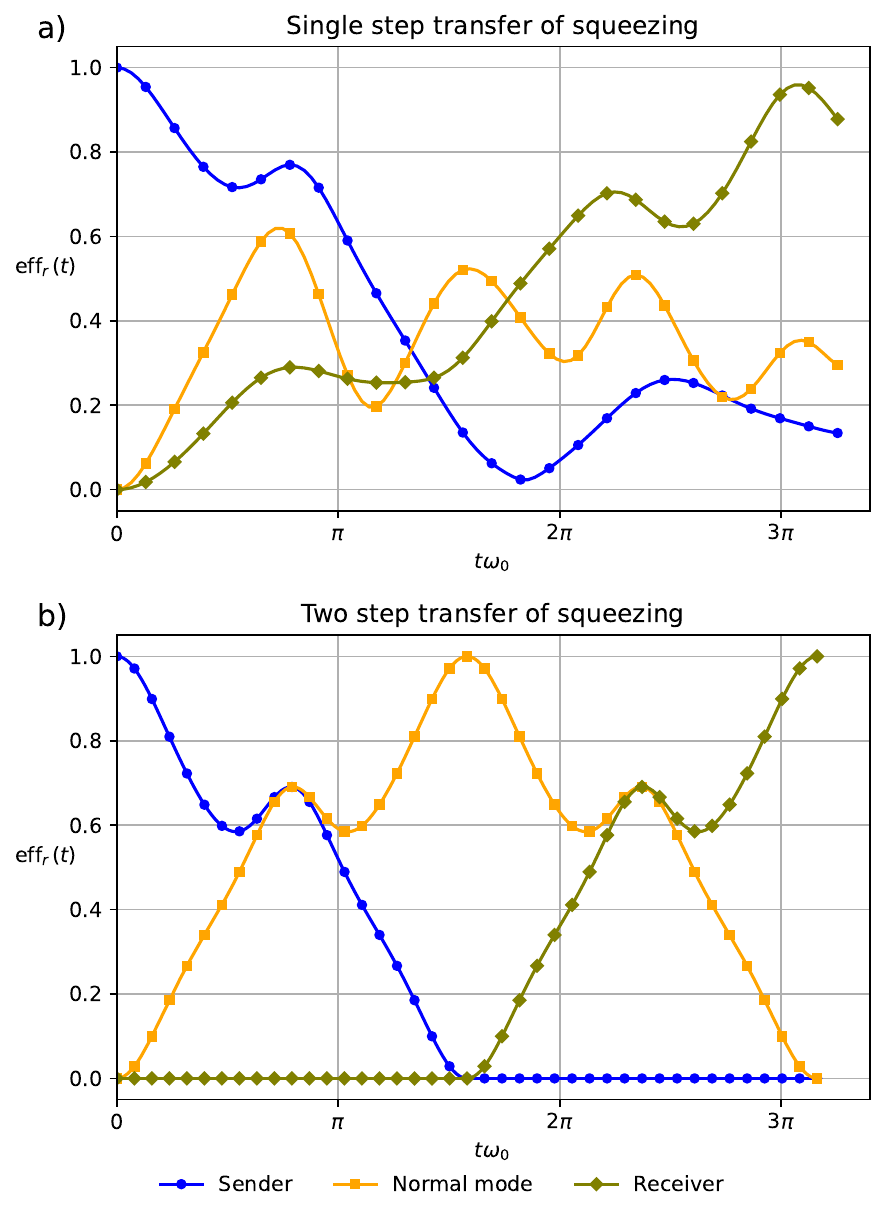}
\caption{\label{fig:idealtransfer} Comparison of \textit{a)} single and \textit{b)} two step transfers in the ideal case where there is only the resonant normal mode at some frequency $\omega_0$. At $t=0$ the sender is in squeezed vacuum with $r=0.5$ whereas the normal mode and the receiver are both in vacuum. In the single step case all oscillators are coupled, resulting in an approximate two-way transfer as the states of the sender and the receiver are approximately swapped. In the two step case only two oscillators are coupled at a time, leading to a perfect one-way transfer. Both protocols can be carried out with different speeds; here top speeds are compared.
}
\end{figure}

State transfer may be achieved by using a normal mode of the network by engineering the effective Hamiltonian to be a chain of just three oscillators: the sender, the normal mode and the receiver. This is a good approximation when the sender and receiver have tuned their frequencies to be resonant with that of the normal mode and their couplings to be weak. More generally, in such chains perfect transfer is known to be possible only for two oscillators \cite{plenio2004dynamics}; we will return to this point shortly. In the case at hand the Hamiltonian of Eq.~\eqref{eq:Hamiltonian} can be diagonalized with some orthogonal matrix $\mathbf{K}$ such that the normal mode position operators $\mathbf{Q}^\top=\{Q_1,Q_2,\ldots,Q_N\}$ become $\mathbf{Q}=\mathbf{K}^\top\mathbf{q}$. Therefore $H_I=-kq_Sq_i=-kq_S\sum_j\mathbf{K}_{ij}Q_i$, which for generic networks means that the sender (and receiver) couple to every normal mode, and in particular to the mode chosen for the transfer. Then that coupling is directly proportional to $k$ and can be freely tuned with it, facilitating engineering of the effective Hamiltonian.

\begin{figure}[h!]
\centering
\includegraphics[trim=0cm 0cm 0.0cm 0cm,clip=true,width=0.48\textwidth]{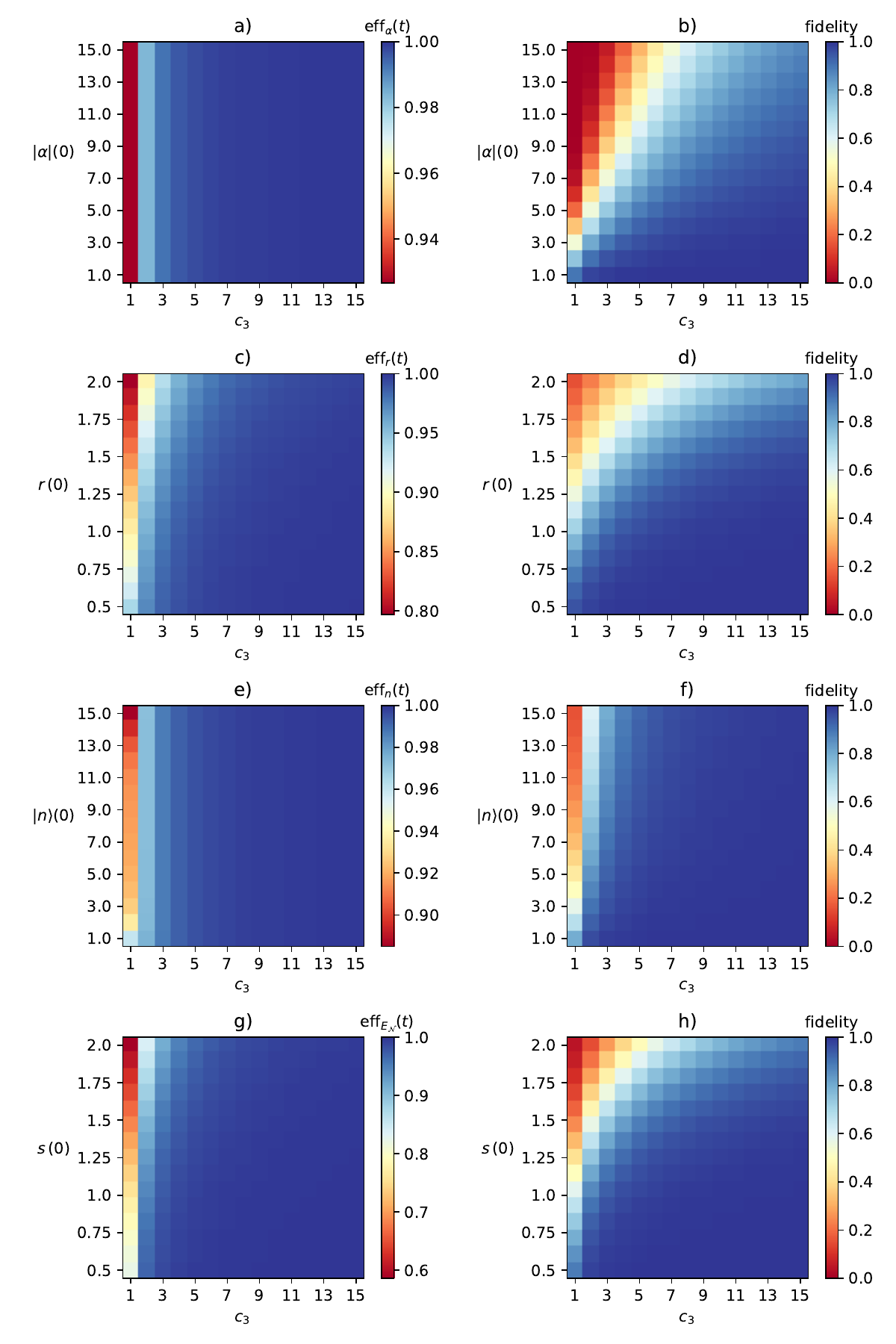}
\caption{\label{fig:fidelityscaling} 
Comparison of transfer performance for different states and transfer speeds for the single step protocol in the ideal case of only a single normal mode. Transfer time increases linearly with the integer $c_3\geq 1$ but performance improves. Efficiency is shown in panels $a)$, $c)$, $e)$ and $g)$ and fidelity in the others. 
Panels $a)$ - $f)$ consider transfers of single mode states:
a coherent state with magnitude of displacement $|\alpha|$ in $a)$, $b)$, 
a squeezed vacuum with squeezing parameter $r$ in  $c)$, $d)$, 
and a number state $|n\rangle$ in $e)$, $f)$. Notably for the number state, which is non-Gaussian, the behavior is similar to the Gaussian states.
A twin beam state with two-mode squeezing parameter $s$ is also considered in $g)$, $h)$, such that only one of the systems couples to the chain; shown are the fraction of transferred entanglement quantified by logarithmic negativity $E_\mathcal{N}$ in $g)$ and fidelity with a twin beam state with the same $s$ but orthogonal phase in $h)$.}
\end{figure}


We consider two inequivalent ways to transfer the state, accomplished by judiciously tuning the effective Hamiltonian and interaction times as explained in App.~\ref{app:swaps}. Detection or state tomography is not considered as it has no effect on the transfer fidelity. In the two step protocol only the sender couples to the network first, and swaps states with the normal mode. Next only the receiver couples to the network and also swaps states with the normal mode. The state is transferred in principle perfectly to the receiver, whereas the original states of the normal mode and the receiver end up in the sender and the normal mode, respectively. In the single step protocol both the sender and the receiver couple to the network simultaneously and for the same duration such that their states are swapped while the normal mode state remains invariant. Both protocols are parameterized by a positive integer such that higher values decrease the coupling strengths but increase transfer times as explained in App.~\ref{app:swaps}, namely $c_2$ and $c_3$ according to whether the transfer involves two coupled systems at a time or all three. On the one hand the single step protocol only approximately achieves the transfer even in the ideal case where the other normal modes are not present, but on the other hand can also be used for two-way transfer as the external systems can both simultaneously send and receive quantum information. We remark that achieving two-way transfer with the first protocol would require an additional third step where the sender again couples to the network to once more swap states with the normal mode. For now, we focus on the ideal case where the effect of non-resonant normal modes vanish, addressing the role of the network structure in later Sections.

The two protocols are compared in Fig.~\ref{fig:idealtransfer}. The state to be transferred is $\Ket{r=0.5,\varphi=0}$ and the figure of merit is the efficiency of squeezing transfer  $\mathrm{eff}_r(t)$. The differences become apparent during the evolution, as does the approximate nature of the single step protocol. In particular, although the transfer time for it should be such that $t\omega_0=3\pi$ as suggested by Eq.~\eqref{eq:appendix3chain}, it can be observed that in practice the efficiency reaches its maximum value a bit later. This suggests a simple optimization strategy for the transfer time for this protocol which will be used later in Sec.~\ref{sec:random}. Although the single step protocol is in principle slightly faster we show in App.~\ref{app:asymptotic} that asymptotically the transfer times are identical whereas the coupling strength required by the two step protocol is stronger by a factor of $\sqrt{2}$.

We now move on to comparing the effect of different states and figures of merit when using the single step protocol for different values of the parameter $c_3$ controlling the couplings and transfer times. Results are shown in Fig.~\ref{fig:fidelityscaling}. In all cases the initial phase is zero. For $\Ket{s,\phi}$, it is the initial state of the sender and an ancillary system which remains uncoupled from all other systems; transferring the state of the sender achieves the transfer of entanglement to be between the receiver and the ancillary system while shifting the phase to be orthogonal with the original $\phi$. High efficiency is easily achieved for displacement, squeezing and entanglement as quantified by logarithmic negativity. High fidelity is more difficult to achieve due to its sensitivity to small phase differences which increases with the magnitude of displacement or the single or two-mode squeezing parameter, however in all cases fidelity continues to improve with $c_3$.

We also consider transfer of a number state in Fig.~\ref{fig:fidelityscaling}\textit{e)}, \textit{f)} as an example of a non-Gaussian state. As noted in App.~\ref{app:swaps}, in the two step protocol the Gaussian nature of the transferred state is irrelevant to the success of the transfer; this seems to extend to the one step protocol as seen from the the transfer efficiencies' behavior. However, compared to the Gaussian states a high fidelity is easier to achieve for the number states due to them being phase invariant. The numerical simulations for the number states are performed differently from the Gaussian states and the details regarding the simulations are included in App.~\ref{app:simulation}.

More generally, some speed should be traded for better performance in both protocols to reduce the detrimental effect of the non-resonant normal modes. Ideally, the normal mode used for transfer should either be relatively far in frequency from the others or at least interact much more strongly with both the sender and the receiver than any nearby modes. The conventional choice is to use the center-of-mass mode, or the normal mode with the lowest frequency $\omega_0$, because the corresponding column in matrix $\mathbf{K}$ is proportional to the unit vector---ensuring decent interaction strength for any pair---and also because of the previously mentioned spectral gap. This leads to favorable scaling of performance with coupling strength and therefore transfer time. In principle, the other modes could lead to faster transfer due to their higher frequencies but typically neither sufficient interaction strength nor separation from other normal modes can be ensured. Cases where at least some of them are useful for transfer are of particular interest as otherwise routing cannot be achieved: only one transfer can take place at a time.

\section{Erd{\H o}s-R{\'e}nyi and modular random networks}\label{sec:random}

In this section we present the routing capabilities of Erd{\H o}s-R{\'e}nyi (ER) and modular random networks when the single step transfer protocol is used, finding that the degree of connection nodes matters to transfer fidelity, and that the community structure of modular networks allows transfer over frequencies other than the-center-of mass mode. Although in the ideal case the two step protocol allows perfect state transfer, it is more vulnerable to the presence of additional normal modes due to the stronger coupling it requires, as shown in App.~\ref{app:asymptotic}. In the general case, the performance difference is in favor of the single step protocol.

\subsection{Slowest normal mode in Erd{\H o}s-R{\'e}nyi random networks}\label{sec:erdosrenyi}

\begin{figure}[ht]
\centering
\includegraphics[trim=0cm 0cm 0.0cm 0cm,clip=true,width=0.48\textwidth]{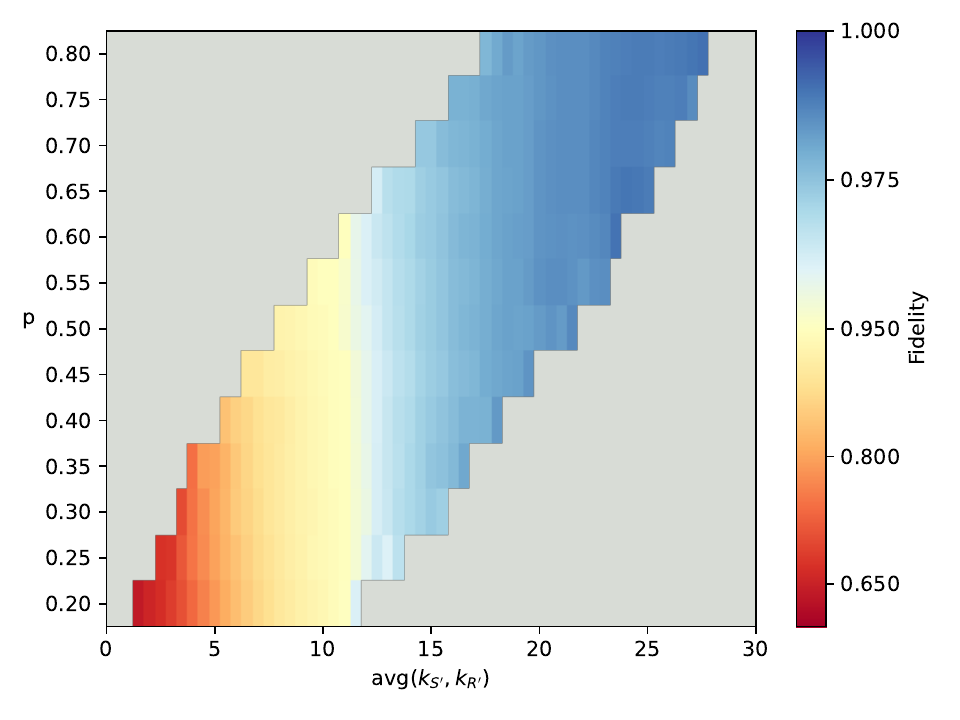}
\caption{\label{fig:er_avgdeg} 
Transfer fidelity of a squeezed vacuum with $r = 1.0$ over an ER network with varying edge probabilities $p$, averaged over sender-receiver pairs where the averages of the degrees of the nodes that the sender and receiver couple to, $k_{S'}$ and $k_{R'}$ respectively, are similar. Note that the colour scale is composed of two linear ranges above and below $F=0.95$ to capture the features of the distributions.}
\end{figure}

We studied how the degree of the nodes to which the external users couple to affects the transfer fidelity of a squeezed vacuum state $\Ket{r=1,\varphi=0}$ in different ER-networks when communicating over the slowest normal mode. For each chosen edge probability $p$ an ensemble of 50 networks of 30 oscillators was generated, and in each network all possible pairs of nodes $(S', R')$ were considered as connection points for the sender and receiver. 

The oscillators have unit mass and frequency, and the coupling strength between network oscillators is 1. The external oscillators are coupled to their respective network nodes with strength $g' = g / \sqrt{N}$, where $g$ corresponds to Eq.~\eqref{eq:appendix3chain} with $c_3 = 7$ and $N=30$ is the number of oscillators. This equates to an effective coupling strength of $g$ to the slowest normal mode. Similarly for the transfer time the value from the ideal case $t_{ideal}$ was used. However, due to the idealized assumptions in the derivation of the transfer time, the maximum fidelity is not reached exactly at the calculated time but slightly after due to an unpredictable phase difference. In addition, the difference between the times varies between realizations. The system was thus simulated for time window $[t_{ideal}, t_{ideal} + 4]$ to capture a full period of the oscillator and thus at least one instant where the phases match. The maximum fidelity within this window was taken as the transfer fidelity.

In order to quantify the importance of both connection nodes' degree, we consider the transfer fidelity as a function of their average degree, $k_{avg} = (k_{S'}+k_{R'})/2$. In each ensemble all pairs with the same $k_{avg}$ were grouped together. The mean of the transfer fidelities achieved between such pairs are presented in Fig.~\ref{fig:er_avgdeg} for each ensemble.
We observe that a higher average degree of the connection nodes translates to better transfer fidelity, while the overall degree distribution, and consequently the spectral gap, of the network does not affect the fidelity. The degree of a node increases its effective frequency, which in turn weakens the coupling to lower frequency normal modes. Higher degree nodes thus couple more weakly to the normal modes closer to the center-of-mass mode, leading to smaller loss of information to the other normal modes. The importance of the spectral gap is also highlighted. In networks with small spectral gaps---such as rings and paths---it must be accounted for with a weak coupling \cite{plenio2005high}, but when the gap is above some threshold, its size does not affect the fidelity.

\begin{figure}[h]
\centering
\includegraphics[trim=0cm 0cm 0.0cm 0cm,clip=true,width=0.48\textwidth]{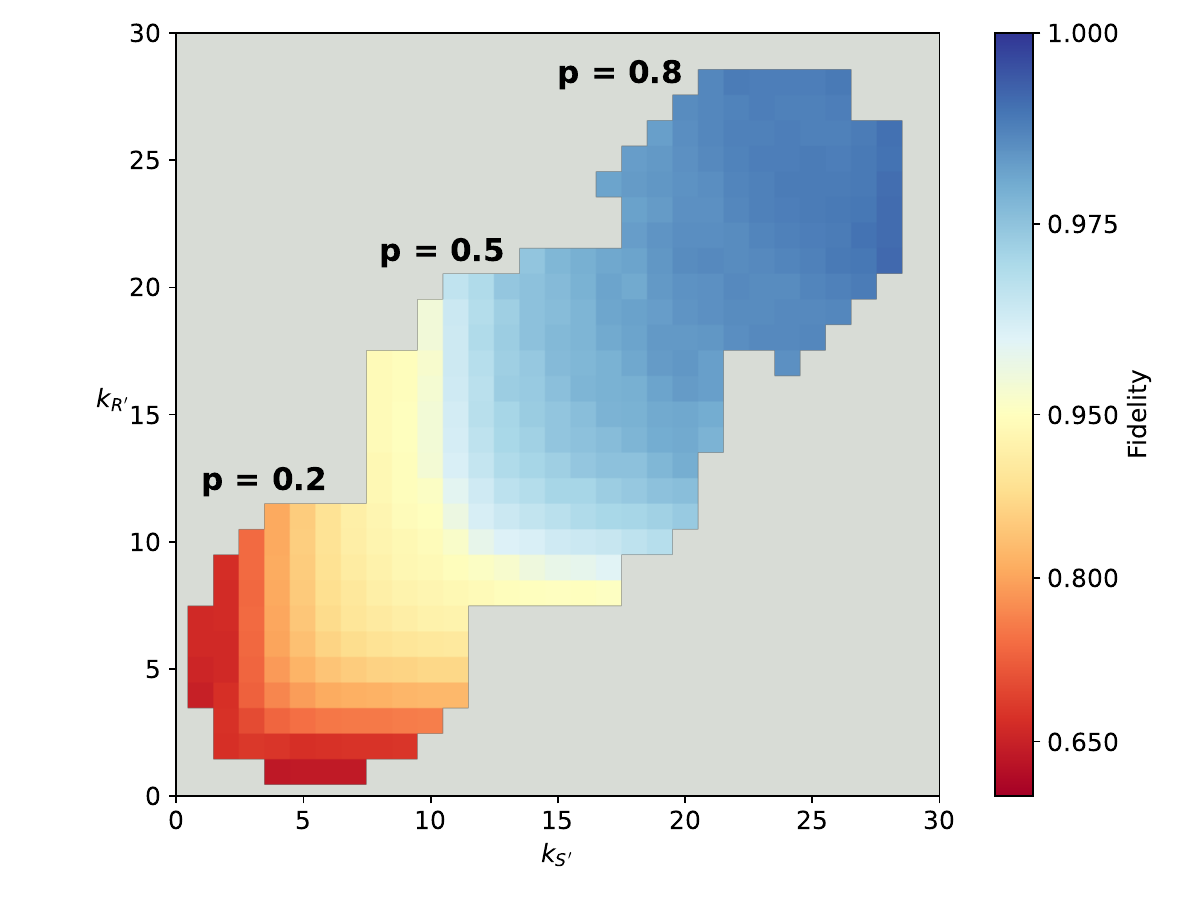}
\caption{\label{fig:er_bothdegs} 
Transfer fidelity of a squeezed vacuum, in each ensemble averaged over pairs with the same degrees on both senders and receivers connection nodes $S'$ and $R'$. Ensembles with $p = 0.2$, $0.5$ and $0.8$ are presented. For lower degree nodes a bottleneck effect is visible: the connection node with lower degree limits the transfer fidelity for the pair. Due to the Poissonian degree distribution of ER-networks, pairs in the middle of each ensembles mass are more common than those at the perimeter. Note that the colour scale is again composed of two linear ranges above and below $F=0.95$.}
\end{figure}

Figure~\ref{fig:er_bothdegs} shows the transfer fidelity as a function of both the degree of the sender and the receiver, for three ensembles with $p = 0.2$, $0.5$ and $0.8$. For pairs in which one of the connection nodes has a low degree, we see that the transfer fidelity is limited by the capability of the weaker node. 

\subsection{Normal modes in modular random networks} \label{sec:modularrandom}

Here we consider modular networks, or networks with community structure. These networks are characterized by high link density within some groups of nodes, and a lower link density between the groups. Such networks may allow transfer over normal modes other than the center of mass mode. In general, a network with $M$ communities has, including the center of mass mode, $M$ normal modes which are well separated from the rest and to which multiple nodes couple with significant strength \cite{chauhan2009spectral}. The coupling strengths to these normal modes depend strongly on the community the node is in. State transfer is then possible within a community over those modes to which the nodes couple strongly, and between communities if the nodes in both communities couple strongly to the same normal mode. If the communities have no good fast normal modes in common, they may in any case communicate over the slowest normal mode.

\begin{figure}
    \centering
    \includegraphics[trim=0cm 0cm 0.0cm 0cm,clip=true,width=0.48\textwidth]{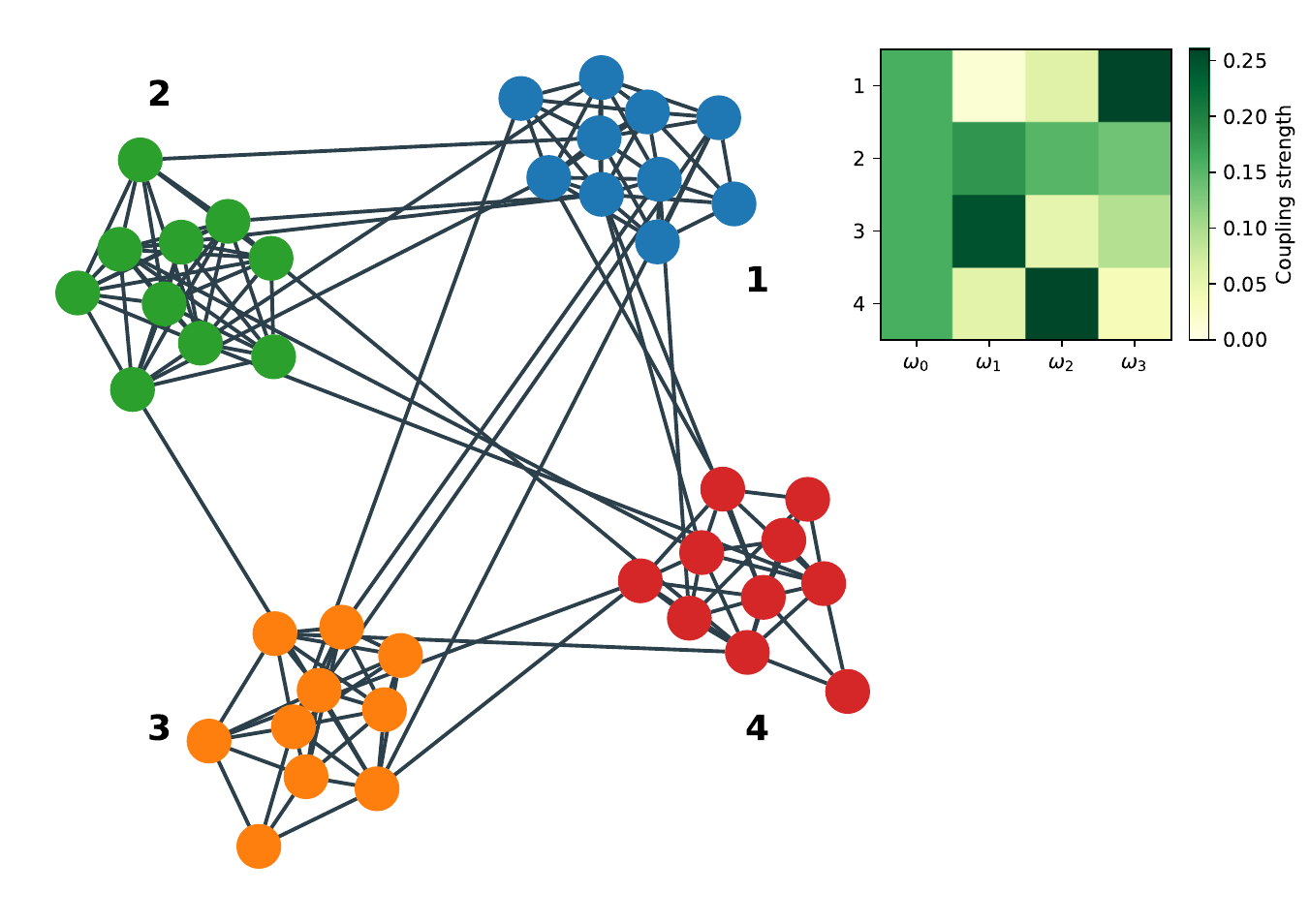}
    \caption{A single realization of a 40-node stochastic block model network, partitioned into four communities of 10. In the inset the mean coupling strengths of nodes in a given community to the four slowest normal modes are shown. Within each community the variance of coupling strengths is small.}
    \label{fig:modular_example}
\end{figure}

To generate modular random networks we used a stochastic block model (SBM) \cite{holland1983stochastic}. In effect the model constructs $M$ communities of ER-networks, and randomly adds edges between the communities. We present here results for a 40-node network of four equally sized communities, with edge probability $p_{\text{w}}=0.75$ within each community and $p_{\text{b}} = 0.025$ between communities. 
These probabilities were chosen to produce a pronounced community structure, with nodes having a high degree from links internal to their community in line with the results of the previous section.
A single example realization is shown in Fig.~\ref{fig:modular_example}, with the mean coupling strengths of the nodes in each community to the four lowest normal modes shown in the inset. As noted earlier, by construction all nodes couple to the center of mass mode with equal strength. For the other modes the coupling strength is strongly dependent on the community, and at least two communities have a significant coupling to each normal mode. We observe that this behaviour holds for most networks with the chosen parameters.

The oscillators have same properties as in Sec.~\ref{sec:erdosrenyi}, and the external oscillators' effective coupling strength is according to Eq.~\eqref{eq:appendix3chain} with $c_3 = 50$. The large value for $c_3$, corresponding to a very weak coupling, was chosen to highlight the difference between good and bad normal modes; with a strong coupling, even good normal modes could allow only low fidelity transfer. The state to be transferred is again the squeezed vacuum $\Ket{r=1,\varphi=0}$. 

\begin{figure}
\centering
\includegraphics[trim=0cm 0cm 0.0cm 0cm,clip=true,width=0.48\textwidth]{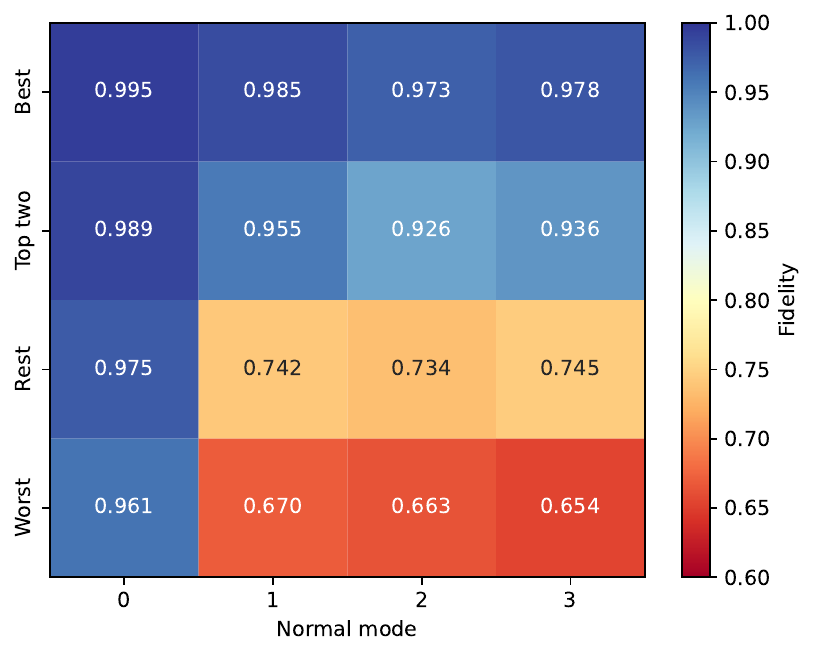}
\caption{\label{fig:modular_fidelity} 
Transfer fidelity of a squeezed vacuum $\Ket{r=1,\varphi=0}$ when communicating over the four slowest normal modes, averaged over 200 realisations of the stochastic block model network. 
In each realization, the mean fidelity within each community over each normal mode was calculated, and the communities were ranked based on their fidelities. \textit{Best} is the fidelity in the community in which the mean fidelity is highest, and correspondingly \textit{worst} is the fidelity of the worst performing community. \textit{Top two} is the mean fidelity within the subgraph consisting of the two best performing communities, accounting for transfers both inside and between the communities.
\textit{Rest} is then the mean transfer fidelity between all other pairs of nodes, with one or both being outside the two best performing communities. It should be noted that the ranking of the communities is not the same for all of the normal modes in each network.
}
\end{figure}

Figure~\ref{fig:modular_fidelity} shows the mean transfer fidelities over the lowest four normal modes for an ensemble of 200 random modular networks. As expected, over the center of mass mode transfer between any pair in the network reaches high fidelity. Over the higher normal modes there is in general a subgraph of only two communities in which high fidelity transfer is possible, while fidelities are low between pairs where one or both participants do not couple to this subgraph. In particular, high fidelity in the `Top two' row indicates the possibility to communicate both within and between different communities. The communities which reach high fidelities may be identified as the communities which couple strongly to the normal mode.

\section{Empirical networks\label{sec:empirical}}

\subsection{Role of community structure}

\begin{figure}[ht]
\centering
\includegraphics[trim=0cm 0cm 0.0cm 0cm,clip=true,width=0.48\textwidth]{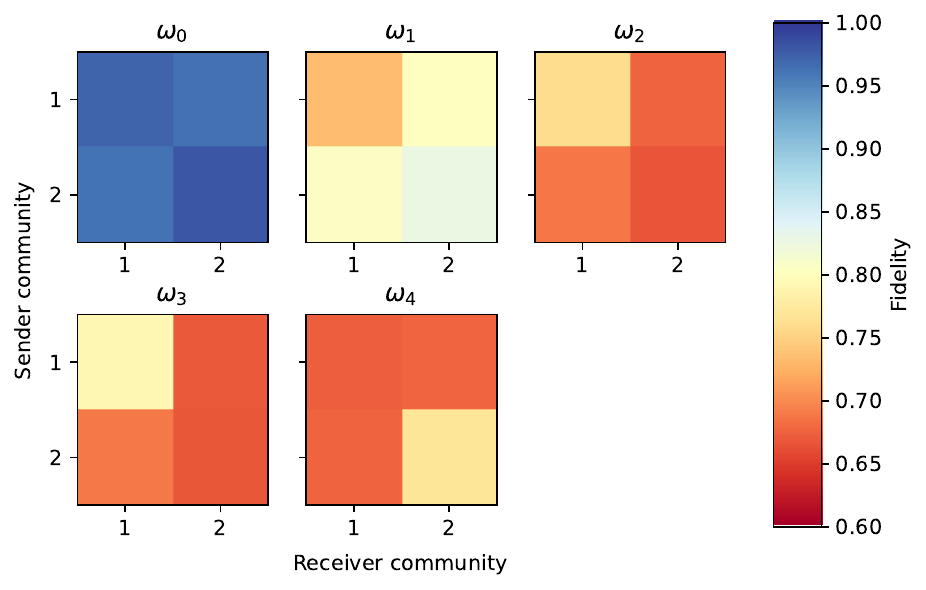}
\caption{\label{fig:karatefids} 
Mean transfer fidelity for a squeezed vacuum $\Ket{r=1,\varphi=0}$ between the original communities of the Karate club network. The fidelities are presented for the five slowest normal modes, from $\omega_0$ to $\omega_4$.}
\end{figure}

\begin{figure}[ht]
\centering
\includegraphics[trim=0cm 0cm 0.0cm 0cm,clip=true,width=0.48\textwidth]{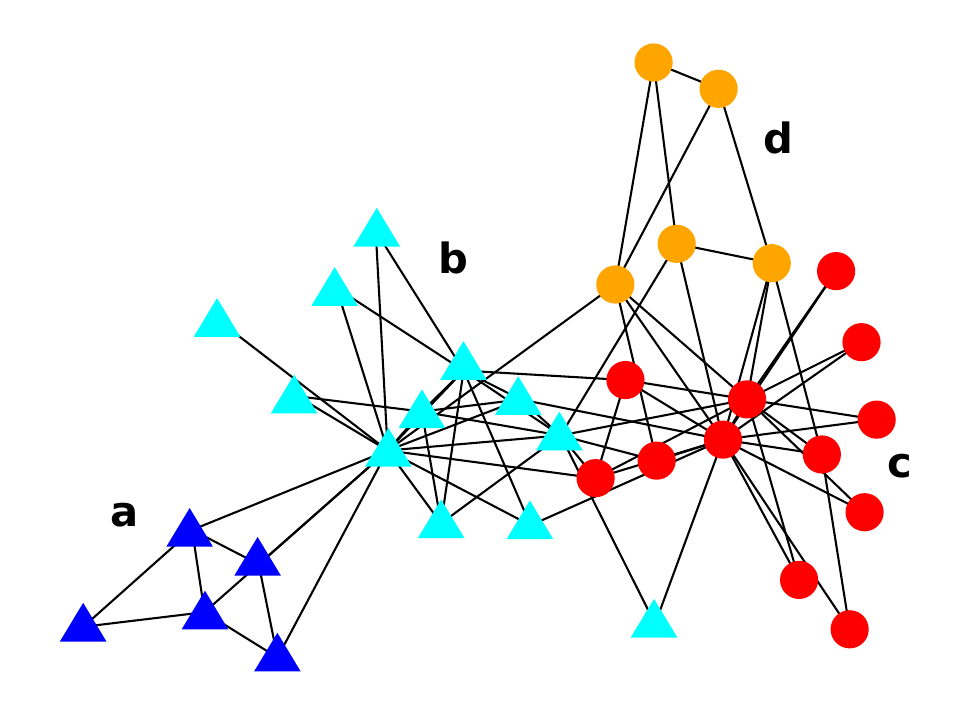}
\caption{\label{fig:karate_communities} 
In the Karate club network, originally consisting of two communities marked here with triangles and circles, further community structure can be found by considering the transfer fidelities. Here we identify one possible partition, indicated by the different colors and letters \textbf{a - d}. Fidelities for this partition are shown in Fig.~\ref{fig:karatefids_fourcomms}.}
\end{figure}

\begin{figure}[ht]
\centering
\includegraphics[trim=0cm 0cm 0.0cm 0cm,clip=true,width=0.48\textwidth]{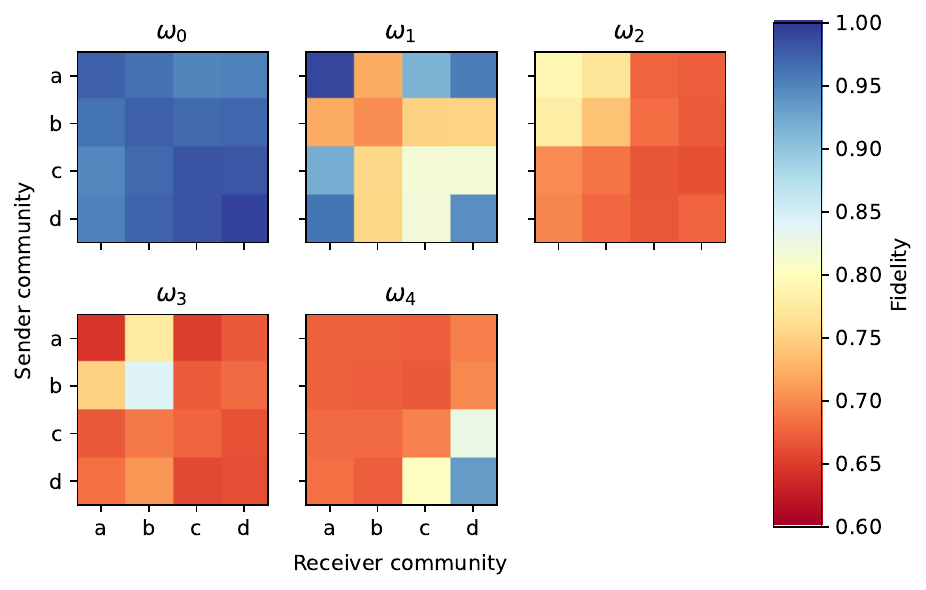}
\caption{\label{fig:karatefids_fourcomms} 
Mean transfer fidelities for the alternate communities \textit{a-d} of the Karate club network shown in Fig.~\ref{fig:karate_communities}. While the fidelities within the original communities are not particularly high, as shown in Fig.~\ref{fig:karatefids}, we may identify these subcommunities within which higher transfer fidelities are achieved. For example, the average fidelity of community 1 over $\omega_1$ is quite low, but in its subcommunity \textit{a} fidelities are close to unity. }
\end{figure}

As an example of a real complex network with community structure we studied the routing properties of Zachary's Karate club network \cite{zachary1977information}, a paradigmatic network with community structure. The network with its division to communities is shown in Fig.~\ref{fig:karate_communities}. A network of QHOs with the unweighted karate club structure was constructed with the same parameters as in Sec.~\ref{sec:modularrandom}, and  transfer of a squeezed vacuum $\Ket{r=1,\varphi=0}$ was again considered. Figure~\ref{fig:karatefids} shows the mean transfer fidelities within and between the communities for the five slowest normal modes. In contrast to the random modular networks in Sec.~\ref{sec:modularrandom}, which had a more pronounced community structure, we find that while communication between the two communities is only feasible over the two slowest normal modes, both communities have additional normal modes over which at least some communication is possible. For even higher normal modes there are in general only a few or no pairs which reach any reasonable fidelity.

Averaging over the communities hides the distribution of transfer fidelities within each community. If we examine the pairwise fidelities separately, we see that which pairs can communicate over a given frequency does not directly correspond to the original communities. We may consider grouping the nodes based on the transfer fidelities, such that all pairs within a group achieve a high fidelity over some frequency. Partitioning the network this way is not unambiguous: some nodes may be able to communicate over multiple frequencies, so that they could be assigned to multiple communities. Conversely some nodes only allow high fidelity transfer over the center of mass mode and thus would not be included in any fidelity-based groups. In Fig.~\ref{fig:karate_communities} we present one possible partition, where both original communities are further split into two, assigning ambiguous nodes arbitrarily. The mean transfer fidelities for these communities are shown in Fig.~\ref{fig:karatefids_fourcomms}.

This partition achieves relatively high modularity $Q$ (computed numerically with the NetworkX Python library \cite{SciPyProceedings_11}), which is a measure of how well the community structure of a network matches a given partition. For the original two community partition $Q\approx 0.36$, while for the fidelity inspired partition in Fig.~\ref{fig:karate_communities} we get $Q \approx 0.41$, indicating that the network is better described with these four communities than the original two.

\begin{figure}[ht]
\centering
\includegraphics[trim=0cm 0cm 0.0cm 1cm,clip=true,width=0.48\textwidth]{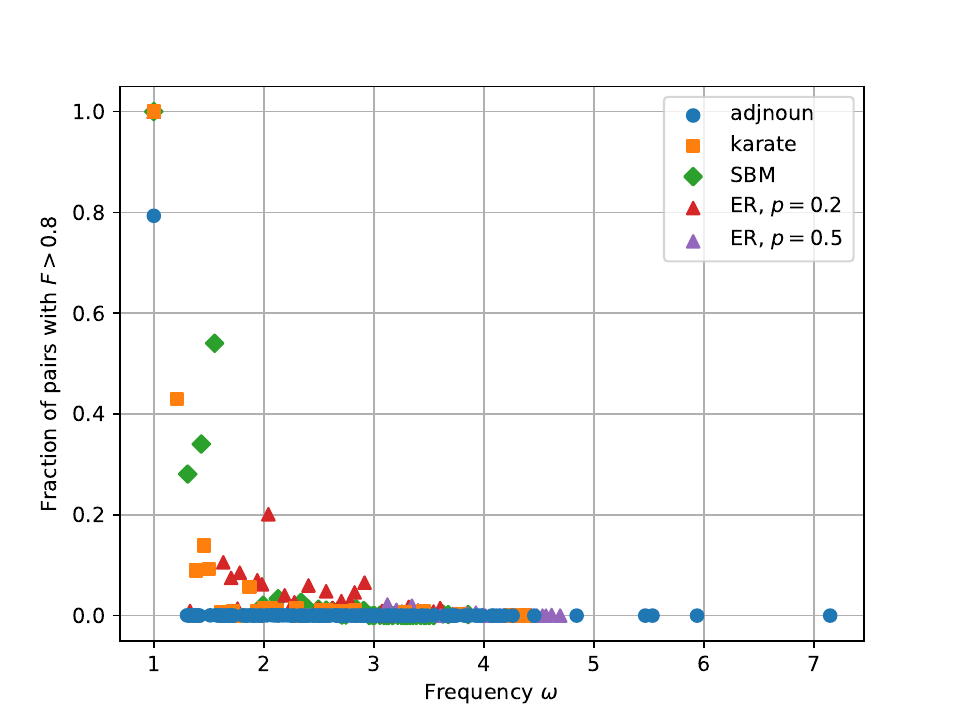}
\caption{\label{fig:good_transfers} 
Fraction of pairs for which $F>0.8$ for all normal modes of the networks presented in the article. The networks have the same parameters as in Sec.~\ref{sec:modularrandom}. For the random networks only a single realization is presented here. We see that the modular networks have a handful of decent normal modes, while the adjective-noun and ER networks mainly allow transfer over the center of mass mode.}
\end{figure}

As another example of a different empirical network we consider the adjective-noun network \cite{newman2006finding}, which is a bipartite network. The network does not have any discernible community structure, and thus we do not expect there to be additional normal modes suitable for routing. The networks are compared in Fig.~\ref{fig:good_transfers}, where the fraction of pairs of nodes which allow transfer fidelity of $F>0.8$ is presented. The random networks are included for reference. We see that the adjective-noun network mainly allows transfer over the center of mass mode, and even then not all pairs reach high fidelities. For all the higher frequency normal modes only a negligible fraction succeeds in high fidelity transfer.

Figure~\ref{fig:good_transfers} highlights another feature of networks that is important for routing: the adjective-noun network has a very dense frequency distribution, while the modular networks have some separation between the suitable normal modes. Although the spectrum of the adjective-noun network has large gaps at the highest frequencies, only few nodes couple to those frequencies, limiting their usefulness for routing. 


\subsection{Comparison with randomized counterparts\label{sec:benchmark}}

Do empirical networks have subtle features not captured by random networks that could give them an edge in state transfer and routing? Here we explore this question using two tentative quantities taking into account both transfer fidelity and time for all normal modes. While they reveal new behavior and properties which warrant further investigation, they have their limitations as from previous results we know that most considered fidelities are low---this is not the full story however, considering for example the results of Fig.~\ref{fig:fidelityscaling}. This motivates the search of more sophisticated measures. Specifically, here we use a coupling strength and transfer time (without optimization) according to Eq.~\eqref{eq:appendix3chain} with $c_3 = 20$, the transfer of a displaced state $\Ket{\alpha=0.75}$ and use as network parameters $\omega_0=0.25$ and $g=0.1$. 

The first quantity attempts to capture the importance of a network node as a point of contact. Let $F_{i,j}^n$ be the transfer fidelity when external users couple to network nodes $i$ and $j$ and transfer happens through the $n$-th normal mode and let $t^n$ be the associated transfer time, inversely proportional to the normal mode frequency but independent of $i$ and $j$. Consider the quantity $C_{i,j}:=\sum_{n=1}^N F_{i,j}^n/t^n$, where $N$ is the network size. Notice that $C_{i,j}=C_{j,i}$. To focus on the importance of a single network node, we consider node capacity $C_i$, defined as
\begin{equation}
   C_{i}:=\sum_{j=1}^N C_{i,j}.
\end{equation}
The value of $C_i$ for some node $i$ is relatively high if it can support higher fidelity transfers at faster speeds compared to the other nodes. To compare different networks' capacities, these values should be then normalized. Here we average the capacities by the network size, $C_i/N$, however one could also consider averaging both $C_{i,j}$ and $C_i$ so that the overall result would be $C_i/N^2$.

\begin{figure}[ht]
\centering
\includegraphics[trim=0cm 0cm 0.0cm 0cm,clip=true,width=0.48\textwidth]{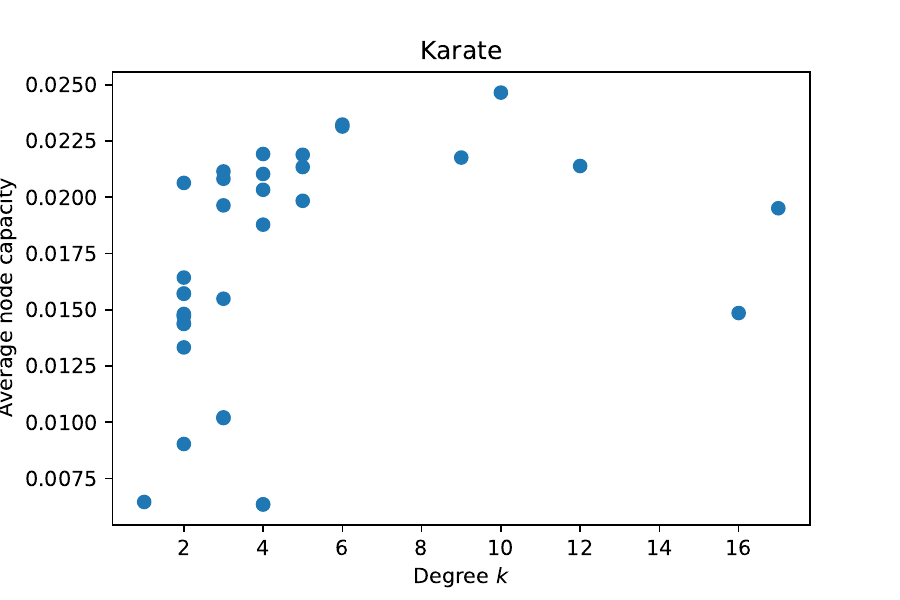}
\caption{\label{fig:nodecapacitykarate} The node capacity averaged by the network size $C_{i}/N$ against the degree of the network node $i$ in the karate club network.}
\end{figure}

\begin{figure}[ht]
\centering
\includegraphics[trim=0cm 0cm 0.0cm 0cm,clip=true,width=0.48\textwidth]{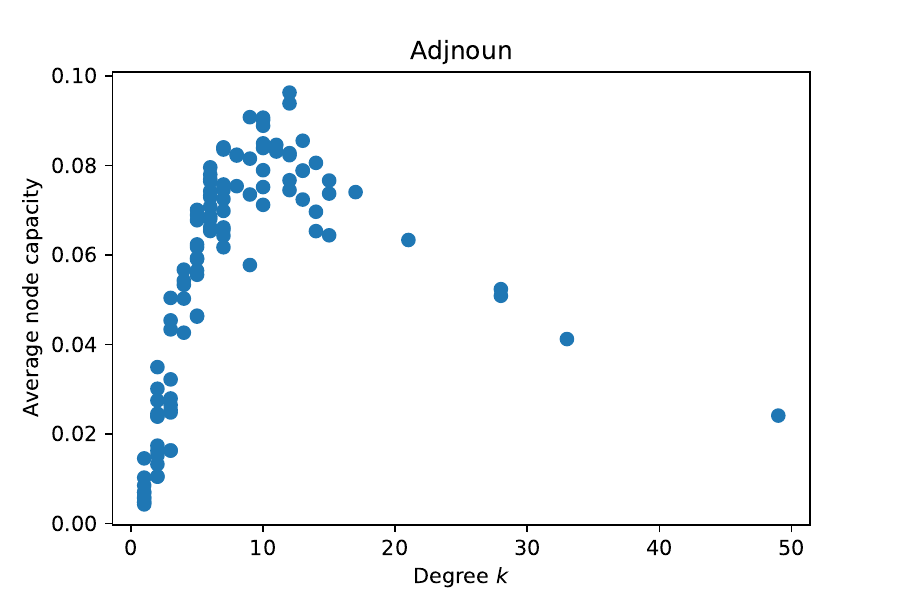}
\caption{\label{fig:nodecapacityadjnoun} The node capacity averaged by the network size $C_{i}/N$ against the degree of the network node $i$ in the adjective--noun network.}
\end{figure}

Results for the two networks are shown in Figs.~\ref{fig:nodecapacitykarate} and \ref{fig:nodecapacityadjnoun}. Like previously a low degree is found to be detrimental, however we also observe a new behavior where nodes with a particularly high degree reach lower node capacities than nodes with an intermediate degree. Overall the behavior can be expected to be robust to changes in network structure as it is observed in two quite different networks, although the effect is particularly strong in the adjective-noun network. Considering results of Sec.~\ref{sec:erdosrenyi}, the nodes with the highest degree can be expected to support high fidelity transfer mostly with the slowest normal mode while nodes with intermediate degree might have a more even spread of transfer fidelities across the available normal modes.

The second considered quantity attempts to capture the maximal throughput of a network by considering the best performing external user pair for each normal mode. Namely, we consider
\begin{equation}
    C_{max}:=\sum_{n=1}^N\mathrm{max}_{i,j}F_{i,j}^n/t^n.
\end{equation}
While the best pairs are not necessarily free of overlapping users, especially for larger networks we expect $C_{max}$ to be a good approximation of the optimal set of transfers that can happen simultaneously. We compare this quantity to that achieved with randomized counterparts for both networks. The random counterparts preserve some but not all features of the original networks as explained in App.~\ref{app:randomization}.

\begin{figure}[h!]
\centering
\includegraphics[trim=0cm 0cm 0.0cm 0cm,clip=true,width=0.45\textwidth]{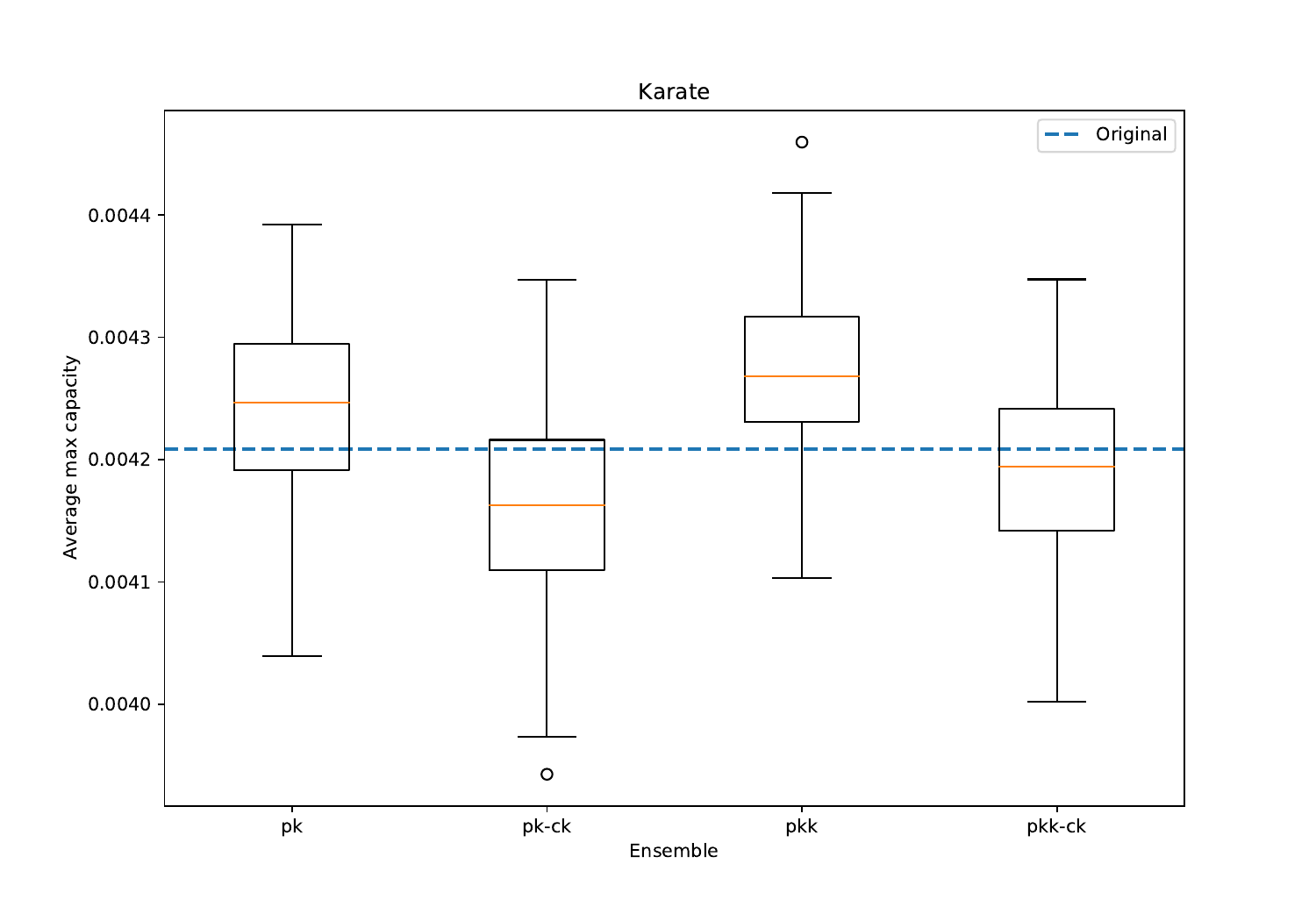}
\caption{\label{fig:maxcapacitykarate} Comparison of $C_{max}/N$ of the karate club network (horizontal dashed line) with its randomized variants (boxes and whiskers). The randomisations preserve some statistical properties of the network. In the pk variants, only the degree distribution is preserved. In pk-ck, also the clustering spectrum, and in pkk, also the degree-degree correlations. In pkk-ck, all three properties are held fixed. }
\end{figure}

\begin{figure}[h!]
\centering
\includegraphics[trim=0cm 0cm 0.0cm 0cm,clip=true,width=0.45\textwidth]{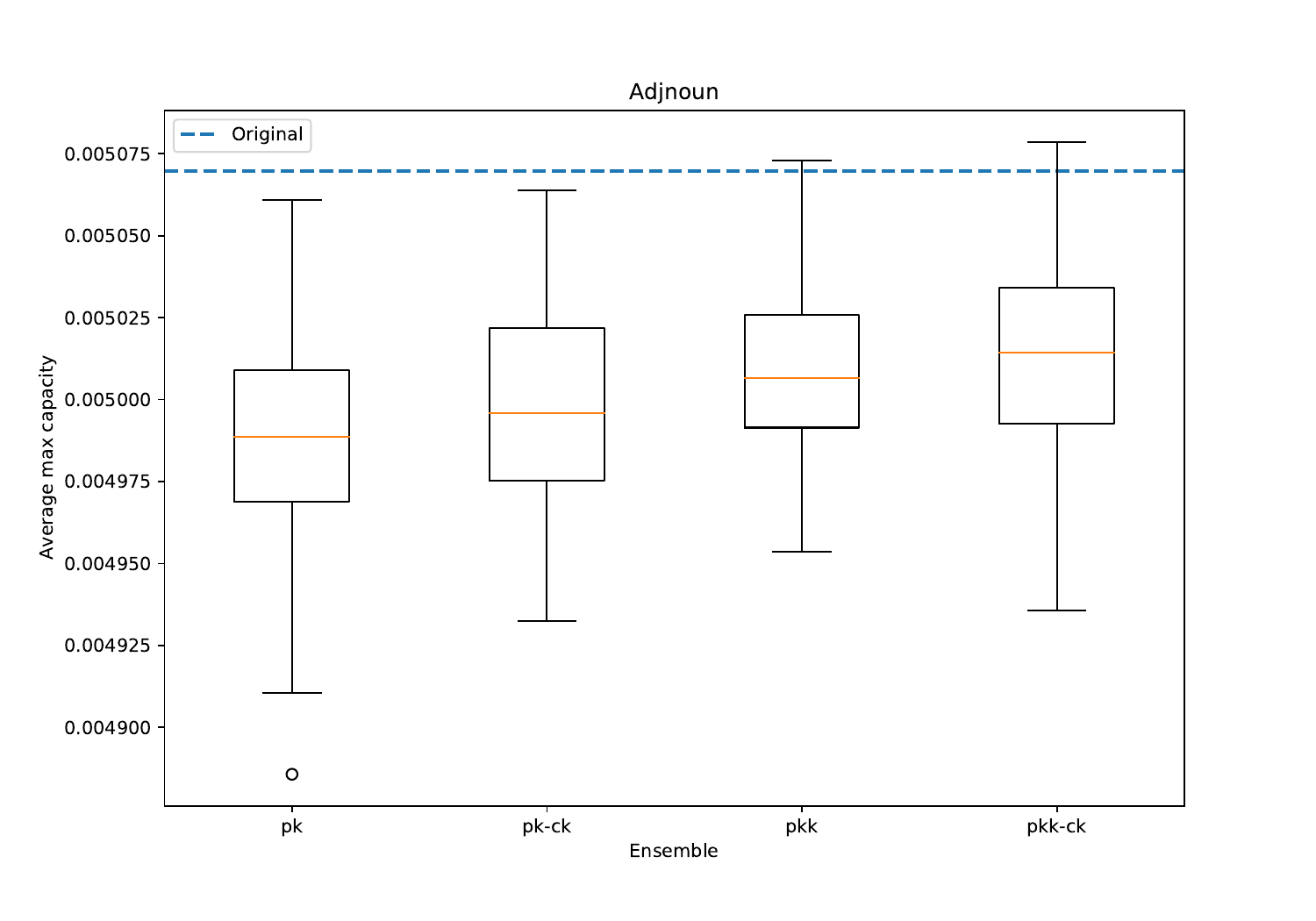}
\caption{\label{fig:maxcapacityadjnoun} Comparison of $C_{max}/N$, defined as in the caption of Fig.~\ref{fig:maxcapacitykarate}, of the adjective--noun network (horizontal dashed line) with its randomized variants (boxes and whiskers).}
\end{figure}

Results are shown in Figs.~\ref{fig:maxcapacitykarate} and \ref{fig:maxcapacityadjnoun}. In the case of karate club the original tends to outperform the random variants if the clustering spectrum $c(k)$ is included in the preserved quantities, but otherwise the roles are reversed. In terms of maximum capacity it would then seem that the network does not reach a particularly high value. Adjective-noun network leads to a very different situation however, with the original outperforming the vast majority of the $400$ random variants. This suggests that the complexity that leads to this advantage is not captured by any of the preserved features. The role of the network size and which features are the important ones is an interesting avenue of further study.

\section{Conclusions}\label{sec:conclusions}

Conventionally, state transfer is considered in either networks with a special structure that can facilitate fast transfer but not between arbitrary nodes \cite{kay2011basics} or transfer over a single normal mode, with a few notable exceptions \cite{wojcik2007multiuser,pemberton2011perfect,paganelli2013routing}. Here we consider multi-user transfer and go beyond the previous work along three major lines: i) exploring the role of the nodes users directly couple to in ER networks, ii) identifying the community structure as the source of multiple useful normal modes in both random and empirical networks, and iii) finally explore the importance of complex structure -- found in empirical networks --  by benchmarking them against their randomized counterparts. In the latter case we observe a clear benefit in the adjective-noun network.

By revealing how community structure can have both nontrivial and beneficial effects we invite the broader research community to also start exploring its role in state transfer, routing and other closely related tasks carried out over networks described by a Hamiltonian. More specifically, our results pave the way for using both complex and generic networks to facilitate routing with local control only, perhaps between multiple nearby quantum processors or even inside a quantum repeater \cite{razavi2018introduction,wei2022towards} embedded in a large scale quantum communication network. Although the experimental realization is a challenge, one could consider, e.g., optomechanical networks \cite{de2016quantum,li2017quantum} or provided that quantum regime can be reached in them, networks of nanomechanical \cite{fon2017complex} or nanoelectromechanical \cite{matheny2019exotic} oscillators. Meanwhile networks with a decent size and otherwise arbitrary structure can currently be created in a multimode quantum optics platform \cite{nokkala2018reconfigurable,renault2023experimental} to provide a testbed for routing. 

We leave open the question of which other features of the network may be relevant to routing, and how properties such as network diameter, degree distribution or clustering affect the transfer performance. In particular, if one can construct an arbitrary network, how should it be designed to maximize routing performance? When is a network suitable only for sequential high fidelity transfer better than a network that can support relatively high fidelity transfer over multiple normal modes, and vice versa?
To systematically explore this and the impact of the previously mentioned network properties calls for further work on quantifying the capacity of a network for both state transfer and for routing. Ideally, it should have a clear physically motivated definition, produce a single real number---to facilitate ranking the networks---with clear interpretation and be applicable to a variety of physical systems beyond just continuous variable Gaussian states. It could be expected to single out networks with many useful normal modes. The tentative quantities we have proposed in Sec.~\ref{sec:benchmark} fulfill some of these requirements already, however especially in the case of Gaussian states there seems to be no obvious way to make the capacity state independent. Furthermore, the ability to trade speed for improved fidelity should be taken into account in some appropriate manner. We hope to tackle this challenge in the future.

The modular structure of our model makes state transfer robust both to static disorder in the frequency of a node and to loss of random links \cite{hahto2024state}. What may cause poor transfer fidelity is weak coupling between the chosen normal mode and the sender or receiver. Here the prospects of noise having a positive role can be considered. This is a well-known phenomenon in excitation transport where it is known as noise-assisted transport \cite{plenio2008dephasing,caruso2009highly,viciani2015observation} or environment-assisted transport \cite{rebentrost2009environment,biggerstaff2016enhancing}, but there are sporadic results concerning quantum state transfer as well \cite{rafiee2013noise,wang2019autonomous}. In the case at hand, balancing a weak coupling to a normal mode leads to a strong coupling in the network basis which can cause the off-resonant normal modes to disturb the transfer. Since it is known that time dependent modulation of the couplings to the network can realize a spectral filter blocking exchange of quantum information to certain frequencies \cite{zwick2014optimized}, it is conceivable that judiciously engineered dynamical noise---that is to say, random time dependent variation with a specific distribution---in these couplings could improve transfer fidelity in such situations. A different approach would be to consider an open network, which would introduce environment-induced decoherence. Since its effect on transfer fidelity hasn't received much attention, investigating it would be an interesting avenue of further research. 

Finally, our results might also serve as a basis to design new quantum inspired community detection algorithms as has been previously done with quantum walks \cite{faccin2014community}, but taking into account explicitly the impact of a heterogeneous link density on the normal mode structure. In general, such algorithms are complementary to the classical methods and may reveal alternative structures.

\section*{Acknowledgements} 
M.H. acknowledges financial support from Vilho, Yrjö and Kalle Väisälä Foundation.
J.N. acknowledges financial support from the Turku Collegium for Science, Medicine and Technology as well as the Academy of Finland under project no. 348854. 

\bibliographystyle{quantum}
\bibliography{references}

\begin{thebibliography}{10}

\bibitem{bose2003quantum}
Sougato Bose.
\newblock ``Quantum communication through an unmodulated spin chain''.
\newblock \href{https://dx.doi.org/10.1103/PhysRevLett.91.207901}{Physical review letters {\bf 91}, 207901}~(2003).

\bibitem{nikolopoulos2014quantum}
Georgios~M Nikolopoulos, Igor Jex, et~al.
\newblock ``Quantum state transfer and network engineering''.
\newblock \href{https://dx.doi.org/10.1007/978-3-642-39937-4}{Springer}. ~(2014).

\bibitem{lewis2023low}
Dylan Lewis, Jo{\~a}o~P Moutinho, Antonio~T Costa, Yasser Omar, and Sougato Bose.
\newblock ``Low-dissipation data bus via coherent quantum dynamics''.
\newblock \href{https://dx.doi.org/10.1103/PhysRevB.108.075405}{Physical Review B {\bf 108}, 075405}~(2023).

\bibitem{kay2011basics}
Alastair Kay.
\newblock ``Basics of perfect communication through quantum networks''.
\newblock \href{https://dx.doi.org/10.1103/PhysRevA.84.022337}{Physical Review A {\bf 84}, 022337}~(2011).

\bibitem{godsil2012state}
Chris Godsil.
\newblock ``State transfer on graphs''.
\newblock \href{https://dx.doi.org/10.1016/j.disc.2011.06.032}{Discrete Mathematics {\bf 312}, 129--147}~(2012).

\bibitem{christandl2005perfect}
Matthias Christandl, Nilanjana Datta, Tony~C Dorlas, Artur Ekert, Alastair Kay, and Andrew~J Landahl.
\newblock ``Perfect transfer of arbitrary states in quantum spin networks''.
\newblock \href{https://dx.doi.org/10.1103/PhysRevA.71.032312}{Physical Review A {\bf 71}, 032312}~(2005).

\bibitem{kostak2007perfect}
V~Kostak, GM~Nikolopoulos, and I~Jex.
\newblock ``Perfect state transfer in networks of arbitrary topology and coupling configuration''.
\newblock \href{https://dx.doi.org/10.1103/PhysRevA.75.042319}{Physical Review A {\bf 75}, 042319}~(2007).

\bibitem{portes2013perfect}
D~Portes, Hilario Rodrigues, Sergio~B Duarte, and Basilio Baseia.
\newblock ``Perfect transfer of quantum states in a network of harmonic oscillators''.
\newblock \href{https://dx.doi.org/10.1140/epjd/e2013-40161-y}{The European Physical Journal D {\bf 67}, 1--6}~(2013).

\bibitem{plenio2005high}
Martin~B Plenio and Fernando~L Semiao.
\newblock ``High efficiency transfer of quantum information and multiparticle entanglement generation in translation-invariant quantum chains''.
\newblock \href{https://dx.doi.org/10.1088/1367-2630/7/1/073}{New Journal of Physics {\bf 7}, 73}~(2005).

\bibitem{wojcik2007multiuser}
Antoni Wojcik, Tomasz {\L}uczak, Pawe{\l} Kurzy{\'n}ski, Andrzej Grudka, Tomasz Gdala, and Ma{\l}gorzata Bednarska.
\newblock ``Multiuser quantum communication networks''.
\newblock \href{https://dx.doi.org/10.1103/PhysRevA.75.022330}{Physical Review A {\bf 75}, 022330}~(2007).

\bibitem{paganelli2013routing}
Simone Paganelli, Salvatore Lorenzo, Tony~JG Apollaro, Francesco Plastina, and Gian~Luca Giorgi.
\newblock ``Routing quantum information in spin chains''.
\newblock \href{https://dx.doi.org/10.1103/PhysRevA.87.062309}{Physical Review A {\bf 87}, 062309}~(2013).

\bibitem{nicacio2016coupled}
F~Nicacio and FL~Semi{\~a}o.
\newblock ``Coupled harmonic systems as quantum buses in thermal environments''.
\newblock \href{https://dx.doi.org/10.1088/1751-8113/49/37/375303}{Journal of Physics A: Mathematical and Theoretical {\bf 49}, 375303}~(2016).

\bibitem{merris1998laplacian}
Russell Merris.
\newblock ``Laplacian graph eigenvectors''.
\newblock \href{https://dx.doi.org/10.1016/S0024-3795(97)10080-5}{Linear algebra and its applications {\bf 278}, 221--236}~(1998).

\bibitem{jamakovic2008robustness}
A~Jamakovic and Piet Van~Mieghem.
\newblock ``On the robustness of complex networks by using the algebraic connectivity''.
\newblock In NETWORKING 2008 Ad Hoc and Sensor Networks, Wireless Networks, Next Generation Internet: 7th International IFIP-TC6 Networking Conference Singapore, May 5-9, 2008 Proceedings 7.
\newblock \href{https://dx.doi.org/10.1007/978-3-540-79549-0_16}{Pages 183--194}.
\newblock Springer~(2008).

\bibitem{jonckheere2015information}
Edmond Jonckheere, Frank~C Langbein, and Sophie~G Schirmer.
\newblock ``Information transfer fidelity in spin networks and ring-based quantum routers''.
\newblock \href{https://dx.doi.org/10.1007/s11128-015-1136-4}{Quantum Information Processing {\bf 14}, 4751--4785}~(2015).

\bibitem{alsulami2022unitary}
Abdulsalam~H Alsulami, Irene D'Amico, Marta~P Estarellas, and Timothy~P Spiller.
\newblock ``Unitary design of quantum spin networks for robust routing, entanglement generation, and phase sensing''.
\newblock \href{https://dx.doi.org/10.1002/qute.202200013}{Advanced Quantum Technologies {\bf 5}, 2200013}~(2022).

\bibitem{plenio2004dynamics}
MB~Plenio, J~Hartley, and Jens Eisert.
\newblock ``Dynamics and manipulation of entanglement in coupled harmonic systems with many degrees of freedom''.
\newblock \href{https://dx.doi.org/10.1088/1367-2630/6/1/036}{New Journal of Physics {\bf 6}, 36}~(2004).

\bibitem{chudzicki2010parallel}
Christopher Chudzicki and Frederick~W Strauch.
\newblock ``Parallel state transfer and efficient quantum routing on quantum networks''.
\newblock \href{https://dx.doi.org/10.1103/PhysRevLett.105.260501}{Physical review letters {\bf 105}, 260501}~(2010).

\bibitem{brown2011coupled}
Kenton~R Brown, Christian Ospelkaus, Yves Colombe, Andrew~C Wilson, Dietrich Leibfried, and David~J Wineland.
\newblock ``Coupled quantized mechanical oscillators''.
\newblock \href{https://dx.doi.org/10.1038/nature09721}{Nature {\bf 471}, 196--199}~(2011).

\bibitem{braunstein1998teleportation}
Samuel~L Braunstein and H~Jeff Kimble.
\newblock ``Teleportation of continuous quantum variables''.
\newblock \href{https://dx.doi.org/10.1103/PhysRevLett.80.869}{Physical Review Letters {\bf 80}, 869}~(1998).

\bibitem{pirandola2006quantum}
Stefano Pirandola and Stefano Mancini.
\newblock ``Quantum teleportation with continuous variables: A survey''.
\newblock \href{https://dx.doi.org/10.1134/S1054660X06100057}{Laser Physics {\bf 16}, 1418--1438}~(2006).

\bibitem{sadlier2020state}
Ronald~J. Sadlier and Travis~S. Humble.
\newblock ``State-dependent routing dynamics in noisy quantum computing devices''~(2021).
\newblock  \href{http://arxiv.org/abs/2012.13131}{arXiv:2012.13131}.

\bibitem{webber2020efficient}
Mark Webber, Steven Herbert, Sebastian Weidt, and Winfried~K Hensinger.
\newblock ``Efficient qubit routing for a globally connected trapped ion quantum computer''.
\newblock \href{https://dx.doi.org/10.1002/qute.202000027}{Advanced Quantum Technologies {\bf 3}, 2000027}~(2020).

\bibitem{bapat2021quantum}
Aniruddha Bapat, Andrew~M Childs, Alexey~V Gorshkov, Samuel King, Eddie Schoute, and Hrishee Shastri.
\newblock ``Quantum routing with fast reversals''.
\newblock \href{https://dx.doi.org/10.22331/q-2021-08-31-533}{Quantum {\bf 5}, 533}~(2021).

\bibitem{sinha2022qubit}
Animesh Sinha, Utkarsh Azad, and Harjinder Singh.
\newblock ``Qubit routing using graph neural network aided monte carlo tree search''.
\newblock \href{https://dx.doi.org/10.1609/aaai.v36i9.21231}{Proceedings of the AAAI Conference on Artificial Intelligence {\bf 36}, 9935--9943}~(2022).

\bibitem{zhan2014perfect}
Xiang Zhan, Hao Qin, Zhi-hao Bian, Jian Li, and Peng Xue.
\newblock ``Perfect state transfer and efficient quantum routing: A discrete-time quantum-walk approach''.
\newblock \href{https://dx.doi.org/10.1103/PhysRevA.90.012331}{Physical Review A {\bf 90}, 012331}~(2014).

\bibitem{li2021discrete}
Hengji Li, Jian Li, and Xiubo Chen.
\newblock ``Discrete-time quantum walk approach to high-dimensional quantum state transfer and quantum routing''~(2021).
\newblock  \href{http://arxiv.org/abs/2108.04923}{arXiv:2108.04923}.

\bibitem{gao2023demonstration}
Huixia Gao, Kunkun Wang, Dengke Qu, Quan Lin, and Peng Xue.
\newblock ``Demonstration of a photonic router via quantum walks''.
\newblock \href{https://dx.doi.org/10.1088/1367-2630/acd270}{New Journal of Physics {\bf 25}, 053011}~(2023).

\bibitem{bottarelli2023quantum}
Alberto Bottarelli, Massimo Frigerio, and Matteo~GA Paris.
\newblock ``Quantum routing of information using chiral quantum walks''.
\newblock \href{https://dx.doi.org/10.1116/5.0146805}{AVS Quantum Science {\bf 5}, 025001}~(2023).

\bibitem{palaiodimopoulos2023chiral}
Nikolaos~E Palaiodimopoulos, Simon Ohler, Michael Fleischhauer, and David Petrosyan.
\newblock ``Chiral quantum router with rydberg atoms''.
\newblock \href{https://dx.doi.org/10.1103/PhysRevA.109.032622}{Physical Review A {\bf 109}, 032622}~(2024).

\bibitem{yousefjani2020simultaneous}
Rozhin Yousefjani and Abolfazl Bayat.
\newblock ``Simultaneous multiple-user quantum communication across a spin-chain channel''.
\newblock \href{https://dx.doi.org/10.1103/PhysRevA.102.012418}{Physical Review A {\bf 102}, 012418}~(2020).

\bibitem{pemberton2011perfect}
Peter~J Pemberton-Ross and Alastair Kay.
\newblock ``Perfect quantum routing in regular spin networks''.
\newblock \href{https://dx.doi.org/10.1103/PhysRevLett.106.020503}{Physical review letters {\bf 106}, 020503}~(2011).

\bibitem{girvan2002community}
Michelle Girvan and Mark~EJ Newman.
\newblock ``Community structure in social and biological networks''.
\newblock \href{https://dx.doi.org/10.1073/pnas.122653799}{Proceedings of the national academy of sciences {\bf 99}, 7821--7826}~(2002).

\bibitem{newman2004finding}
Mark~EJ Newman and Michelle Girvan.
\newblock ``Finding and evaluating community structure in networks''.
\newblock \href{https://dx.doi.org/10.1103/PhysRevE.69.026113}{Physical review E {\bf 69}, 026113}~(2004).

\bibitem{fortunato2010community}
Santo Fortunato.
\newblock ``Community detection in graphs''.
\newblock \href{https://dx.doi.org/10.1016/j.physrep.2009.11.002}{Physics reports {\bf 486}, 75--174}~(2010).

\bibitem{tikochinsky1979diagonalization}
Yoel Tikochinsky.
\newblock ``On the diagonalization of the general quadratic hamiltonian for coupled harmonic oscillators''.
\newblock \href{https://dx.doi.org/10.1063/1.524093}{Journal of Mathematical Physics {\bf 20}, 406--408}~(1979).

\bibitem{ferraro2005gaussian}
Alessandro Ferraro, Stefano Olivares, and Matteo~GA Paris.
\newblock ``Gaussian states in quantum information''.
\newblock \href{https://dx.doi.org/10.48550/arXiv.quant-ph/0503237}{Napoli Series on physics and Astrophysics}. Bibliopolis. ~(2005).

\bibitem{adesso2014continuous}
Gerardo Adesso, Sammy Ragy, and Antony~R Lee.
\newblock ``Continuous variable quantum information: Gaussian states and beyond''.
\newblock \href{https://dx.doi.org/10.1142/S1230161214400010}{Open Systems \& Information Dynamics {\bf 21}, 1440001}~(2014).

\bibitem{chauhan2009spectral}
Sanjeev Chauhan, Michelle Girvan, and Edward Ott.
\newblock ``Spectral properties of networks with community structure''.
\newblock \href{https://dx.doi.org/10.1103/PhysRevE.80.056114}{Physical Review E {\bf 80}, 056114}~(2009).

\bibitem{holland1983stochastic}
Paul~W. Holland, Kathryn~Blackmond Laskey, and Samuel Leinhardt.
\newblock ``Stochastic blockmodels: {First} steps''.
\newblock \href{https://dx.doi.org/https://doi.org/10.1016/0378-8733(83)90021-7}{Social Networks {\bf 5}, 109--137}~(1983).

\bibitem{zachary1977information}
Wayne~W Zachary.
\newblock ``An information flow model for conflict and fission in small groups''.
\newblock Journal of anthropological research {\bf 33}, 452--473~(1977).
\newblock  url:~\url{http://www.jstor.org/stable/3629752}.

\bibitem{SciPyProceedings_11}
Aric~A. Hagberg, Daniel~A. Schult, and Pieter~J. Swart.
\newblock ``Exploring network structure, dynamics, and function using {NetworkX}''.
\newblock In Ga\"el Varoquaux, Travis Vaught, and Jarrod Millman, editors, Proceedings of the 7th Python in Science Conference.
\newblock Pages 11 -- 15.
\newblock Pasadena, CA USA~(2008).

\bibitem{newman2006finding}
Mark~EJ Newman.
\newblock ``Finding community structure in networks using the eigenvectors of matrices''.
\newblock \href{https://dx.doi.org/10.1103/PhysRevE.74.036104}{Physical review E {\bf 74}, 036104}~(2006).

\bibitem{razavi2018introduction}
Mohsen Razavi.
\newblock ``An introduction to quantum communications networks: Or, how shall we communicate in the quantum era?''.
\newblock \href{https://dx.doi.org/10.1088/978-1-6817-4653-1}{2053-2571}. Morgan \& Claypool Publishers. ~(2018).

\bibitem{wei2022towards}
Shi-Hai Wei, Bo~Jing, Xue-Ying Zhang, Jin-Yu Liao, Chen-Zhi Yuan, Bo-Yu Fan, Chen Lyu, Dian-Li Zhou, You Wang, Guang-Wei Deng, et~al.
\newblock ``Towards real-world quantum networks: A review''.
\newblock \href{https://dx.doi.org/10.1002/lpor.202100219}{Laser \& Photonics Reviews {\bf 16}, 2100219}~(2022).

\bibitem{de2016quantum}
GD~de~Moraes~Neto, FM~Andrade, V~Montenegro, and S~Bose.
\newblock ``Quantum state transfer in optomechanical arrays''.
\newblock \href{https://dx.doi.org/10.1103/PhysRevA.93.062339}{Physical Review A {\bf 93}, 062339}~(2016).

\bibitem{li2017quantum}
Wenlin Li, Chong Li, and Heshan Song.
\newblock ``Quantum synchronization and quantum state sharing in an irregular complex network''.
\newblock \href{https://dx.doi.org/10.1103/PhysRevE.95.022204}{Physical Review E {\bf 95}, 022204}~(2017).

\bibitem{fon2017complex}
Warren Fon, Matthew~H Matheny, Jarvis Li, Lev Krayzman, Michael~C Cross, Raissa~M D’Souza, James~P Crutchfield, and Michael~L Roukes.
\newblock ``Complex dynamical networks constructed with fully controllable nonlinear nanomechanical oscillators''.
\newblock \href{https://dx.doi.org/10.1021/acs.nanolett.7b02026}{Nano letters {\bf 17}, 5977--5983}~(2017).

\bibitem{matheny2019exotic}
Matthew~H Matheny, Jeffrey Emenheiser, Warren Fon, Airlie Chapman, Anastasiya Salova, Martin Rohden, Jarvis Li, Mathias Hudoba~de Badyn, M{\'a}rton P{\'o}sfai, Leonardo Duenas-Osorio, et~al.
\newblock ``Exotic states in a simple network of nanoelectromechanical oscillators''.
\newblock \href{https://dx.doi.org/10.1126/science.aav7932}{Science {\bf 363}, eaav7932}~(2019).

\bibitem{nokkala2018reconfigurable}
Johannes Nokkala, Francesco Arzani, Fernando Galve, Roberta Zambrini, Sabrina Maniscalco, Jyrki Piilo, Nicolas Treps, and Valentina Parigi.
\newblock ``Reconfigurable optical implementation of quantum complex networks''.
\newblock \href{https://dx.doi.org/10.1088/1367-2630/aabc77}{New Journal of Physics {\bf 20}, 053024}~(2018).

\bibitem{renault2023experimental}
P.~Renault, J.~Nokkala, G.~Roeland, N.Y. Joly, R.~Zambrini, S.~Maniscalco, J.~Piilo, N.~Treps, and V.~Parigi.
\newblock ``Experimental optical simulator of reconfigurable and complex quantum environment''.
\newblock \href{https://dx.doi.org/10.1103/PRXQuantum.4.040310}{PRX Quantum {\bf 4}, 040310}~(2023).

\bibitem{hahto2024state}
Markku Hahto, Jyrki Piilo, and Johannes Nokkala.
\newblock ``State transfer in noisy modular quantum networks''.
\newblock \href{https://dx.doi.org/10.1002/qute.202400316}{Advanced Quantum Technologies {\bf 8}, 2400316}~(2025).

\bibitem{plenio2008dephasing}
Martin~B Plenio and Susana~F Huelga.
\newblock ``Dephasing-assisted transport: quantum networks and biomolecules''.
\newblock \href{https://dx.doi.org/10.1088/1367-2630/10/11/113019}{New Journal of Physics {\bf 10}, 113019}~(2008).

\bibitem{caruso2009highly}
Filippo Caruso, Alex~W Chin, Animesh Datta, Susana~F Huelga, and Martin~B Plenio.
\newblock ``Highly efficient energy excitation transfer in light-harvesting complexes: The fundamental role of noise-assisted transport''.
\newblock \href{https://dx.doi.org/10.1063/1.3223548}{The Journal of Chemical Physics {\bf 131}, 105106}~(2009).

\bibitem{viciani2015observation}
Silvia Viciani, Manuela Lima, Marco Bellini, and Filippo Caruso.
\newblock ``Observation of noise-assisted transport in an all-optical cavity-based network''.
\newblock \href{https://dx.doi.org/10.1103/PhysRevLett.115.083601}{Physical Review Letters {\bf 115}, 083601}~(2015).

\bibitem{rebentrost2009environment}
Patrick Rebentrost, Masoud Mohseni, Ivan Kassal, Seth Lloyd, and Al{\'a}n Aspuru-Guzik.
\newblock ``Environment-assisted quantum transport''.
\newblock \href{https://dx.doi.org/10.1088/1367-2630/11/3/033003}{New Journal of Physics {\bf 11}, 033003}~(2009).

\bibitem{biggerstaff2016enhancing}
Devon~N Biggerstaff, Ren{\'e} Heilmann, Aidan~A Zecevik, Markus Gr{\"a}fe, Matthew~A Broome, Alessandro Fedrizzi, Stefan Nolte, Alexander Szameit, Andrew~G White, and Ivan Kassal.
\newblock ``Enhancing coherent transport in a photonic network using controllable decoherence''.
\newblock \href{https://dx.doi.org/10.1038/ncomms11282}{Nature communications {\bf 7}, 11282}~(2016).

\bibitem{rafiee2013noise}
Morteza Rafiee, Cosmo Lupo, and Stefano Mancini.
\newblock ``Noise to lubricate qubit transfer in a spin network''.
\newblock \href{https://dx.doi.org/10.1103/PhysRevA.88.032325}{Physical Review A—Atomic, Molecular, and Optical Physics {\bf 88}, 032325}~(2013).

\bibitem{wang2019autonomous}
Chen Wang and Jeffrey~M Gertler.
\newblock ``Autonomous quantum state transfer by dissipation engineering''.
\newblock \href{https://dx.doi.org/10.1103/PhysRevResearch.1.033198}{Physical Review Research {\bf 1}, 033198}~(2019).

\bibitem{zwick2014optimized}
Analia Zwick, Gonzalo~A {\'A}lvarez, Guy Bensky, and Gershon Kurizki.
\newblock ``Optimized dynamical control of state transfer through noisy spin chains''.
\newblock \href{https://dx.doi.org/10.1088/1367-2630/16/6/065021}{New Journal of Physics {\bf 16}, 065021}~(2014).

\bibitem{faccin2014community}
Mauro Faccin, Piotr Migda{\l}, Tomi~H Johnson, Ville Bergholm, and Jacob~D Biamonte.
\newblock ``Community detection in quantum complex networks''.
\newblock \href{https://dx.doi.org/10.1103/PhysRevX.4.041012}{Physical Review X {\bf 4}, 041012}~(2014).

\bibitem{nokkala2018quantum}
Johannes Nokkala.
\newblock ``Quantum complex networks''.
\newblock PhD thesis.
\newblock University of Turku.
\newblock ~(2018).

\bibitem{johansson2012qutip}
J~Robert Johansson, Paul~D Nation, and Franco Nori.
\newblock ``{QuTiP}: An open-source python framework for the dynamics of open quantum systems''.
\newblock \href{https://dx.doi.org/10.1016/j.cpc.2012.02.021}{Computer physics communications {\bf 183}, 1760--1772}~(2012).

\bibitem{scutaru1998fidelity}
Horia Scutaru.
\newblock ``Fidelity for displaced squeezed thermal states and the oscillator semigroup''.
\newblock \href{https://dx.doi.org/10.1088/0305-4470/31/15/025}{Journal of Physics A: Mathematical and General {\bf 31}, 3659}~(1998).

\bibitem{banchi2015quantum}
Leonardo Banchi, Samuel~L Braunstein, and Stefano Pirandola.
\newblock ``Quantum fidelity for arbitrary gaussian states''.
\newblock \href{https://dx.doi.org/10.1103/PhysRevLett.115.260501}{Physical review letters {\bf 115}, 260501}~(2015).

\bibitem{adesso2005gaussian}
Gerardo Adesso and Fabrizio Illuminati.
\newblock ``Gaussian measures of entanglement versus negativities: Ordering of two-mode gaussian states''.
\newblock \href{https://dx.doi.org/10.1103/PhysRevA.72.032334}{Physical Review A—Atomic, Molecular, and Optical Physics {\bf 72}, 032334}~(2005).

\bibitem{orsini2015quantifying}
Chiara Orsini, Marija~M Dankulov, Pol Colomer-de Sim{\'o}n, Almerima Jamakovic, Priya Mahadevan, Amin Vahdat, Kevin~E Bassler, Zolt{\'a}n Toroczkai, Mari{\'a}n Bogun{\'a}, Guido Caldarelli, et~al.
\newblock ``Quantifying randomness in real networks''.
\newblock \href{https://dx.doi.org/10.1038/ncomms9627}{Nature communications {\bf 6}, 8627}~(2015).

\bibitem{randnetgen}
Pol Colomer-de Sim\'on.
\newblock ``Randnetgen''.
\newblock \url{https://github.com/polcolomer/RandNetGen}~(2014).

\bibitem{serrano2006clustering}
M~{\'A}ngeles Serrano and Marian Boguna.
\newblock ``Clustering in complex networks. {I.} {General} formalism''.
\newblock \href{https://dx.doi.org/10.1103/PhysRevE.74.056114}{Physical Review E—Statistical, Nonlinear, and Soft Matter Physics {\bf 74}, 056114}~(2006).

\end{thebibliography}

\appendix

\section{Transferred states\label{app:covariancesandvectors}}

Single mode Gaussian states are completely defined in terms of their covariance matrix $\sigma(\mathbf{x})$ and first moments vector $\langle\mathbf{x}\rangle$. For an oscillator of frequency $\omega$, they read
\begin{equation}
\begin{cases}
\sigma(\mathbf{x})=(n_{\text{th}}+\frac{1}{2})\begin{pmatrix}
(y+z_{\text{cos}})\omega^{-1} & z_{\text{sin}} \\
z_{\text{sin}} & (y-z_{\text{cos}})\omega 
\end{pmatrix},\\
\langle\mathbf{x}\rangle=\begin{pmatrix}
\Re(\alpha)\sqrt{2\omega^{-1}} \\
\Im(\alpha)\sqrt{2\omega}
\end{pmatrix},
\end{cases}
\label{eq:singlemodestates}
\end{equation}
where $y=\cosh{(2r)}$, $z_{\text{cos}}=\cos{(\varphi)}\sinh{(2r)}$ and $z_{\text{sin}}=-\sin{(\varphi)}\sinh{(2r)}$, $n_{\text{th}}$ is the mean number of thermal excitations and $\Re(\alpha)$ and $\Im(\alpha)$ are the real and imaginary parts of the displacement, respectively. For squeezed states $\Ket{r,\varphi}$ we set $\alpha=0$ and for coherent states $\Ket{\alpha}$ we set $r=0$.

Two mode squeezed vacuum states $\Ket{s,\phi}$ correspond to a covariance matrix
\begin{equation}
    \sigma(\mathbf{x})=\frac{1}{2}\begin{pmatrix}
    \omega^{-1}y & 0 & \omega^{-1}z_{\text{cos}} & z_{\text{sin}}\\
    0 & \omega y & z_{\text{sin}} & -\omega z_{\text{cos}}\\
   \omega^{-1}z_{\text{cos}} & z_{\text{sin}} & \omega^{-1}y & 0\\
    z_{\text{sin}} & -\omega z_{\text{cos}} & 0 & \omega y
    \end{pmatrix}
\label{eq:twomodesqueezedstates}
\end{equation}
where this time $y=\cosh(2s)$, $z_{\text{cos}}=\cos(\phi)\sinh(2s)$ and $z_{\text{sin}}=\sin(\phi)\sinh(2s)$.

For the non-Gaussian number states we cannot employ the description in terms of $\sigma$ and $\langle \mathbf{x} \rangle$. Instead they are presented in the discrete Fock basis, where each oscillator has a $d$-dimensional Hilbert space with basis vectors $\{\Ket{n}\}_{n=0}^{d}$. In theory $d=\infty$, but due to computational limitations $d<\infty$. The number states correspond to the basis vectors, so that an $n$-photon state is $\Ket{n}$.

\section{State transfer in short oscillator chains with linear couplings\label{app:swaps}}

Consider identical oscillators with frequencies $\omega_0$ and linear couplings of the form $-gq_iq_j$. The Hamiltonian is of the general form $H=\frac{\mathbf{p}^\top\mathbf{p}}{2}+\mathbf{q}^\top(\omega_0^2\mathbf{I}/2-g\mathbf{V}/2)\mathbf{q}$ where $\mathbf{V}$ is the adjacency matrix of the chain. The Hamiltonian can be diagonalized with an orthogonal matrix $\mathbf{K}$ consisting of the eigenvectors of matrix $\omega_0^2\mathbf{I}/2-g\mathbf{V}/2$. The eigenvalues are of the form $\Omega_i^2/2$ where $\Omega_i$ is a normal mode frequency. Suppose the initial form of the operators is $\mathbf{x}(0)^\top=\{q_1(0),q_2(0),\ldots,p_1(0),p_2(0),\ldots\}$. Then its form at some later time $t$ is recovered from $\mathbf{x}(t)=\mathbf{S}(t)\mathbf{x}(0)$ where
\begin{equation}
\mathbf{S}(t)=
    \begin{pmatrix}
\mathbf{K} & \mathbf{0}\\
\mathbf{0} & \mathbf{K}
\end{pmatrix}
\begin{pmatrix}
\mathbf{\Delta}_{\cos}^\Omega & \mathbf{\Delta}_{\mathbf{\Omega}}^{-1}\mathbf{\Delta}_{\sin}^\Omega\\
-\mathbf{\Delta}_{\mathbf{\Omega}}\mathbf{\Delta}_{\sin}^\Omega & \mathbf{\Delta}_{\cos}^\Omega
\end{pmatrix}
\begin{pmatrix}
\mathbf{K} & \mathbf{0}\\
\mathbf{0} & \mathbf{K}
\end{pmatrix}^\top.
\label{eq:appendixHamiltonian}
\end{equation}
Here $\mathbf{\Delta}_{\cos}^\Omega$ and $\mathbf{\Delta}_{\sin}^\Omega$ are diagonal matrices with diagonal elements $\cos(\Omega_i t)$ and $\sin(\Omega_i t)$, respectively, and $\mathbf{\Delta}_{\mathbf{\Omega}}$ is a diagonal matrix with diagonal elements $\Omega_i$.

Permuting the initial states between all oscillators, which in particular leads to state transfer, requires that at some time $t>0$ the matrix $\mathbf{S}(t)$ becomes block diagonal with two coinciding square blocks defining some permutation matrix $\mathbf{P}$. By substituting Eq.~\eqref{eq:appendixHamiltonian} into $\mathbf{S}(t)=\begin{pmatrix}
\mathbf{P} & \mathbf{0}\\
\mathbf{0} & \mathbf{P}
\end{pmatrix}$ and moving the block matrices containing $\mathbf{K}$ to R.H.S., we find the conditions
\begin{equation}    \mathbf{\Delta}_{\cos}^\Omega=\mathbf{K}^\top\mathbf{P}\mathbf{K},\quad\mathbf{\Delta}_{\sin}^\Omega=\mathbf{0}.
    \label{eq:appendixconditions}
\end{equation}
It is worth stressing that if a solution is found then the permutation applies to arbitrary states although we consider only Gaussian states for convenience.

As pointed out in Sec.~\ref{sec:shortchains}, the cases relevant for our purposes concern permutations in a chain of three oscillators. Without a loss of generality we may assume that the normal mode frequency is $\omega_0$ and that the sender and the receiver have tuned their frequencies to match it. Focusing first on applying twice the state swap between two oscillators, $\mathbf{P}=\pm\begin{pmatrix}
0 & 1\\
1 & 0
\end{pmatrix}$, $\mathbf{K}=\frac{1}{\sqrt{2}}\begin{pmatrix}
1 & -1\\
1 & 1
\end{pmatrix}$ and $\mathbf{\Omega}=\{\sqrt{\omega_0^2-g},\sqrt{\omega_0^2+g}\}$. The freedom to use either sign for the matrix $\mathbf{P}$ is because it will be used twice, and the second application will restore the original phase if necessary. The diagonalization is valid when $g<\omega_0^2$. From Eq.~\eqref{eq:appendixconditions} we get $\cos(t\sqrt{\omega_0^2-g})=-\cos(t\sqrt{\omega_0^2+g})=1$ for positive $\mathbf{P}$ and $-1$ otherwise whereas in both cases $\sin(t\sqrt{\omega_0^2-g})=\sin(t\sqrt{\omega_0^2+g})=0$. Taking $\omega_0$ to be fixed, we solve for $g$ and $t$ and find a two-parameter family of solutions $g=\frac{(1+c')^2-c''^2}{(1+c')^2+c''^2}\omega_0^{2}$, $t=\sqrt{\frac{1}{2}+c'(1+c'/2)+c''^2/2}\pi\omega_0^{-1}$ where $c'\geq c''\geq 1$ are integers. Transfer time $t$ increases with $c'$ and $c''$ whereas setting $c''=a c'+b$ gives $\lim_{c'\rightarrow\infty}g=\frac{1-a^2}{1+a^2}$. We are interested in particular in the solutions where $g$ tends to $0$ as transfer time increases as this reduces the detrimental effect of the other normal modes, which corresponds to choosing $c'=c'':=c_2$. Therefore we will use
\begin{equation}
    g=\frac{1+2c_2}{1+2c_2+2c_2^2}\omega_0^2,\quad t=\sqrt{\frac{1}{2}+c_2+c_2^2}\pi\omega_0^{-1}
    \label{eq:appendix2chain}
\end{equation}
where $c_2\geq 1$ is an integer, which leads to an exact application of matrix $\mathbf{P}$. The time for the transfer between sender and receiver is twice the time above.

We turn our attention to finding a single time independent Hamiltonian for achieving the permutation in a chain of three oscillators. Now Eq.~\eqref{eq:appendixconditions} leads to $\cos(t\omega_0)=-1$ and $\cos(t\sqrt{\omega_0^2-\sqrt{2}g})=\cos(t\sqrt{\omega_0^2+\sqrt{2}g})=1$. Solving the first equation for the transfer time gives $t=\frac{(2 c_3+1)\pi}{\omega_0}$ where $c_3\geq 1$ is an integer. The other two conditions cannot be satisfied simultaneously for this $t$, but by requiring that $g$ is as weak as possible for a given value of $c_3$, equivalently for a given transfer time $t$, taking the average of the solutions for $\cos(t\sqrt{\omega_0^2-\sqrt{2}g})=1$ and $\cos(t\sqrt{\omega_0^2+\sqrt{2}g})=1$ gives
\begin{equation}
    g=\frac{\sqrt{2}}{2 c_3+1}\omega_0^2,\quad t=(2 c_3+1)\pi\omega_0^{-1}
    \label{eq:appendix3chain}
\end{equation}
where $c_3\geq 1$ is an integer. This leads to an approximate state swap between the end points of the chain.

\section{Asymptotic behavior of transfer times and coupling strengths\label{app:asymptotic}}

Let $t_2$ and $t_1$ be the times to transfer the state from the sender to the receiver over the normal mode for the two step and single step protocols, respectively, given by Eqs.~\eqref{eq:appendix2chain} and \eqref{eq:appendix3chain}. Let $c_2=c_3=c$; then $c\geq1$ and an integer. Judicious manipulation of the equations shows that
\begin{equation}
t_2=\sqrt{1+1/(2c+1)^{2}}t_1,
\end{equation}
that is to say $t_2$ approaches $t_1$ from above. The approach is rather fast however, as for example for $c=4$ the factor is already less than $1.01$.

Let now $g_2$ and $g_1$ be the corresponding coupling strengths. This time we find that
\begin{equation}
g_2=\left(1-\frac{1}{2+4c(1+c)}\right)\sqrt{2}g_1,
\end{equation}
i.e. $g_2$ approaches $\sqrt{2}g_1$ from below. Like transfer times, also here the values quickly converge with the factor becoming over 0.99 at $c=5$.

\section{Details about numerical simulations\label{app:simulation}}

We begin by describing the case of Gaussian states, for which simulations are based on exact diagonalization and numerical real matrices. We construct the initial covariance matrix of the full system, diagonalize the total Hamiltonian to find the symplectic matrix that propagates it to the final covariance matrix, and pick the elements corresponding to the receiver. The vector of first moments is handled similarly as seen below. In all cases, both fidelity of Eq.~\eqref{eq:fidelity} and efficiency of Eq.~\eqref{eq:efficiency} can be calculated directly from the covariance matrices and first moments vectors of the sender and the receiver.

In more detail, the initial covariance matrices are based on Eqs.~\eqref{eq:singlemodestates} and \eqref{eq:twomodesqueezedstates}. The vacuum is recovered as a special case of the former by setting $n_{\text{th}}=0$ and $r=0$. The initial covariance matrix of the network is a product state of such vacua, except in Sec.~\ref{sec:benchmark} where it is the ground state of the network Hamiltonian (see, e.g., Equation (3.18) of \cite{nokkala2018quantum}). There are no initial correlations between the sender, the receiver and the network. Consequently the full initial covariance matrix is formed simply by a direct sum of the three initial covariance matrices, followed by a reordering according to the chosen order of operators.

Let $\sigma(\mathbf{x}(0))$ be this initial numerical covariance matrix where $\mathbf{x}(0)$ is the column vector of operators at time $t=0$. Then $\sigma(\mathbf{x}(t))=\mathbf{S}(t)\sigma(\mathbf{x}(0))(\mathbf{S}(t))^\top$. For first moments vector $\langle\mathbf{x}(t)\rangle=\mathbf{S}(t)\langle\mathbf{x}(0)\rangle$. The construction of the symplectic matrix $\mathbf{S}(t)$ is essentially described in the first paragraph of App.~\ref{app:swaps}. Specifically, we start from the matrix of Eq.~\eqref{eq:Hamiltonian}, add the terms for the sender and receiver frequencies with a direct sum, and include the linear coupling terms. The resulting matrix is real and symmetric and can be diagonalized by numerically finding its eigendecomposition which then determines $\mathbf{S}(t)$ through Eq.~\eqref{eq:appendixHamiltonian}. The elements of the final covariance matrix and first moments vectors for the receiver are picked from $\sigma(\mathbf{x}(t))$ and $\langle\mathbf{x}(t)\rangle$, respectively.

In Sec.~\ref{sec:model} we also consider the transfer of a number state in the ideal case as an example of a non-Gaussian state transfer. The procedure for their simulation is substantially different from the Gaussian case, utilizing a finite Hilbert space whose dimension is truncated at a suitable number. For our simulations we chose $d=20$: the states considered go up to $\Ket{n = 15}$, and even for the highest photon initial states we found that the probabilities of states with $n > 20$ in the received state were insignificant, below $10^{-6}$.

The Hamiltonian is translated to the basis of creation and annihilation operators, giving 
$H = \omega_0\sum_i (a^\dagger_i a_i + \frac{1}{2}) - \frac{g}{2\omega_0}\sum_{i,j} V_{i,j} (a_i a_j + a_i^\dagger a_j + \mathrm{h.c.})$. Using this Hamiltonian we then numerically solve the Schrödinger equation $\Ket{\Psi(t)} = e^{-iHt}\Ket{\Psi(0)}$ using the QuTiP Python library \cite{johansson2012qutip}. Here $\Ket{\Psi(0)} = \Ket{n}\otimes\Ket{0}\otimes\Ket{0}$ is the initial state where the sender has $n$ photons and the other oscillators are in the vacuum. From the final state $\Ket{\Psi(t)}$ the density matrix of the receiver is obtained by tracing out the other oscillators.

To calculate the fidelity in the single mode Gaussian case we use \cite{scutaru1998fidelity}
\begin{equation}
    \begin{aligned}
        \mathcal{F}(\rho_1,\rho_2) =& \frac{1}{\sqrt{\Delta + \Lambda} - \sqrt{\Lambda}} \times \\
        &e^{-\frac{1}{2}(\langle \mathbf{x}_2 \rangle - \langle \mathbf{x}_1 \rangle)^T (\sigma_1 + \sigma_2)^{-1} (\langle \mathbf{x}_2 \rangle - \langle \mathbf{x}_1 \rangle)},
    \end{aligned}
\end{equation}
where $\Lambda = 4(\mathrm{det}(\sigma_1) - 1/4) (\mathrm{det}(\sigma_2) - 1/4)$ and $\Delta = \mathrm{det}(\sigma_1 + \sigma_2)$, with  $\sigma_1 = \sigma_S(\mathbf{x}(0))$ and $\sigma_2 = \sigma_R(\mathbf{x}(t))$ and similarly for $\langle \mathbf{x}_1 \rangle$ and $\langle \mathbf{x}_2 \rangle$. For the non-Gaussian states the fidelity is straightforward to calculate directly from the density matrices using the definition in Eq.~\eqref{eq:fidelity}. Fidelity needs to be found for multimode Gaussian states in Fig.~\ref{fig:fidelityscaling}$h$. For this, we apply Eqs.~(11)--(15) of \cite{banchi2015quantum} which gives us the numerical value of $\mathrm{Tr}\sqrt{\sqrt{\rho_1}\rho_2\sqrt{\rho_1}}$. We square the result to find $\mathcal{F}(\rho_1,\rho_2)$.

Efficiency requires the final value of various quantities. Displacement $\alpha$ and squeezing parameter $r$ can be solved from the final covariance matrix of the receiver using Eqs.~\eqref{eq:singlemodestates}.
For the number states the expectation value for the number of excitations $\langle n \rangle$ is calculated from the density matrix of the receiver $\rho_R$ with $\langle n \rangle = \mathrm{Tr}(n \rho_R)$, with $n = a^\dagger a$ being the number operator.
Logarithmic negativity is defined to be
\begin{equation}
    \mathrm{E}(\sigma)=\max\{0,-\log_2{2\tilde{d}_{-}}\}.
\label{eq:loganega}
\end{equation}
where $\tilde{d}_{-}$ is the smaller of the two so-called symplectic eigenvalues of a two-mode covariance matrix $\sigma$. Its value is calculated from the elements of $\sigma$ using Eq.~(13) of \cite{adesso2005gaussian}. 

\section{Network randomization algorithms\label{app:randomization}}

Complex networks can exhibit a wealth of non-trivial topological features, from heterogeneous degree distributions to degree-degree correlations and clustering.
In order to assess the impact of these properties on transport efficiency, we constructed, for both real network considered, several ensembles containing 100 networks each in which some of these properties are preserved.
The randomisations were carried out using the code developed for Ref.~\cite{orsini2015quantifying} (repository available at \cite{randnetgen}).

For any input network, the program generates an ensemble of 100 random networks preserving either 
\begin{itemize}
    \item[i.] only the degree distribution $P(k)$,
    \item[ii.] degree distribution $P(k)$ and degree-degree correlations $P(k, k')$,
    \item[iii.] degree distribution $P(k)$ and clustering spectrum $c(k)$ or
    \item[iv.] degree distribution $P(k)$, degree-degree correlations $P(k, k')$, and clustering spectrum $c(k)$
\end{itemize}
of the original network. Here, degree distribution $P(k)$ is the probability distribution of a node having degree $k$, degree-degree correlations (joint degree distribution) $P(k, k')$ are the probabilities of nodes with degrees $k$ and $k'$ having a link between them, and clustering spectrum $c(k)$ is the spectrum of mean local clustering coefficient of nodes of a given degree $k$. Local clustering coefficient of a node is the probability that a pair of its neighbours are also neighbours of each other \cite{serrano2006clustering}.

\end{document}